%% file: main.tex
\pdfoutput=1
\documentclass{article} % For LaTeX2e

\usepackage{color}
\usepackage[dvipsnames,svgnames]{xcolor}

\usepackage{iclr2022_conference,times}

\input{math_commands}

\usepackage{cancel}

\input{macro}

\usepackage{graphicx}
\usepackage{booktabs}

\usepackage{graphbox}
\usepackage[font=small,skip=3pt,subrefformat=parens]{subcaption}
\usepackage{paralist}
\usepackage{multirow}

\usepackage{makecell}
\usepackage{tabularx}
\usepackage{arydshln}

\usepackage{algorithm}
\usepackage{setspace}
\usepackage[noend]{algpseudocode}

\usepackage[normalem]{ulem}

\usepackage[hidelinks,colorlinks,citecolor=NavyBlue]{hyperref}    % remove ugly boxes
\usepackage{url}

\title{Clean Images are Hard to Reblur:\\ Exploiting the Ill-Posed Inverse Task for \\Dynamic Scene Deblurring}

\author{Seungjun Nah, Sanghyun Son, Jaerin Lee \& Kyoung Mu Lee\\
ASRI, Department of Electrical and Computer Engineering\\
Seoul National University\\
Seoul, 08826, Republic of Korea\\
\texttt{seungjun.nah@gmail.com, \{thstkdgus35, ironjr, kyoungmu\}@snu.ac.kr}
}

\author{Seungjun Nah$^{1,2}$\thanks{Most work was done at SNU} , Sanghyun Son$^{1}$, Jaerin Lee$^{1}$ \& Kyoung Mu Lee$^{1}$\\
$^{1}$ASRI, Department of ECE, Seoul National University, Seoul, Korea \quad
$^{2}$NVIDIA\\
\texttt{seungjun.nah@gmail.com, \{thstkdgus35, ironjr, kyoungmu\}@snu.ac.kr}
}

\iclrfinalcopy % Uncomment for camera-ready version, but NOT for submission.
\begin{document}

\maketitle

\input{sections/camera_ready/0_abstract}
\input{sections/camera_ready/1_introduction}

\input{sections/camera_ready/2_related_works}
\input{sections/camera_ready/3_methods}
\input{sections/camera_ready/4_experiments}
\input{sections/camera_ready/5_conclusion}
\input{sections/camera_ready/aa_acknowledgment}

\bibliography{iclr2022_conference}
\bibliographystyle{iclr2022_conference}

\appendix
\input{sections/camera_ready/a_appendix}

\end{document}

%% file: math_commands.tex
\usepackage{amsmath,amsfonts,bm}

\def\eqref#1{equation~\ref{#1}}
\def\1{\bm{1}}

\DeclareMathAlphabet{\mathsfit}{\encodingdefault}{\sfdefault}{m}{sl}
\SetMathAlphabet{\mathsfit}{bold}{\encodingdefault}{\sfdefault}{bx}{n}

%% file: macro.tex
\newcommand{\norm}[1]{\|{#1}\|}
\newcommand{\paren}[1]{\left( #1 \right)}

\newcommand{\figcspace}{\vspace{-2mm}}
\newcommand{\figspace}{\vspace{-4mm}}
\newcommand{\tabcspace}{\vspace{-2mm}}
\newcommand{\tabspace}{\vspace{-4mm}}

%% file: sections/camera_ready/0_abstract.tex
\begin{abstract}

The goal of dynamic scene deblurring is to remove the motion blur in a given image.
Typical learning-based approaches implement their solutions by minimizing the L1 or L2 distance between the output and the reference sharp image.
Recent attempts adopt visual recognition features in training to improve the perceptual quality.
However, those features are primarily designed to capture high-level contexts rather than low-level structures such as blurriness.
Instead, we propose a more direct way to make images sharper by exploiting the inverse task of deblurring, namely, reblurring.
Reblurring amplifies the remaining blur to rebuild the original blur, however, a well-deblurred clean image with zero-magnitude blur is hard to reblur.
Thus, we design two types of reblurring loss functions for better deblurring.
The supervised reblurring loss at training stage compares the amplified blur between the deblurred and the sharp images.
The self-supervised reblurring loss at inference stage inspects if there noticeable blur remains in the deblurred.
Our experimental results on large-scale benchmarks and real images demonstrate the effectiveness of the reblurring losses in improving the perceptual quality of the deblurred images in terms of NIQE and LPIPS scores as well as visual sharpness.

% The goal of dynamic scene deblurring is to remove the motion blur in a given image. Typical learning-based approaches implement their solutions by minimizing the L1 or L2 distance between the output and the reference sharp image. Recent attempts adopt visual recognition features in training to improve the perceptual quality. However, those features are primarily designed to capture high-level contexts rather than low-level structures such as blurriness. Instead, we propose a more direct way to make images sharper by exploiting the inverse task of deblurring, namely, reblurring. Reblurring amplifies the remaining blur to rebuild the original blur, however, a well-deblurred clean image with zero-magnitude blur is hard to reblur. Thus, we design two types of reblurring loss functions for better deblurring. The supervised reblurring loss at training stage compares the amplified blur between the deblurred and the sharp images. The self-supervised reblurring loss at inference stage inspects if noticeable blur remains in the deblurred. Our experimental results on large-scale benchmarks and real images demonstrate the effectiveness of the reblurring losses in improving the perceptual quality of the deblurred images in terms of NIQE and LPIPS scores as well as visual sharpness.

\end{abstract}

%% file: sections/camera_ready/1_introduction.tex
\section{Introduction}
\label{sec:intro}

% Dynamic scene deblurring - ill-posed, prior knowledge used.
% Motion blur is a common photographic artifact that occurs when the camera moves and the scene changes during the exposure in dynamic environments.
% Motion blur is a common photographic artifact that occurs by the camera movement and the scene change during the exposure in dynamic environments.
Motion blur commonly arises when the cameras move or scene changes during the exposure in dynamic environments.
% Dynamic scene deblurring is a challenging ill-posed problem as both the locally-varying blur kernel and the latent sharp image have to be found from a large solution space.
Dynamic scene deblurring is a challenging ill-posed task finding both the locally-varying blur and the latent sharp image from a large solution space.
Traditional approaches~\citep{hirsch2011fast,whyte2012non,Kim_2013_ICCV,Kim_2014_CVPR} tried to alleviate the ill-posedness by using statistical prior on sharp images such as gradient sparsity.

% durring the image capturing process in dynamic environments.
% Dynamic scene deblurring aims to remove such unwanted blur 

% Classical optimization methods - human knowledge in prior
% Dynamic scene deblurring aims to remove unwanted motion blur from an image and recover the latent sharp image.
% Blind image deblurring is a challenging ill-posed problem as both the locally-varying blur kernel and the latent image have to be found from large solution space.
% Traditional optimization-based approaches~\citep{hirsch2011fast,whyte2012non,Kim_2013_ICCV,Kim_2014_CVPR} tried to relieve the ill-posedness by designing priors that reflect the statistical properties of sharp images.%desired solutions.
% With $\mathbf{B}$, $\mathbf{L}$ as vectorized blurry and the latent images and the large kernel matrix $\mathbf{K}$, a typical energy formulation is
% \begin{equation}
%     \argmin_{K,L} \norm{\mathbf{B} - \mathbf{K}\mathbf{L}} + E_\text{prior}\paren {\mathbf{K}, \mathbf{L}}.
%     \label{eq:eq_formulation}
% \end{equation}

\input{sections/figs/kernel}
Instead of using such handcrafted knowledge, recent methods take advantage of large-scale datasets as well as deep neural networks~\citep{Nah_2017_CVPR,Su_2017_CVPR,noroozi2017motion,Nah_2019_CVPR_Workshops_REDS,Shen_2019_ICCV_Human_aware}.
Usually, the learning is driven by minimizing the pixel-wise distance to the ground truth, e.g., L1 or L2, so that the PSNR between the deblurred and the sharp reference can be maximized.
By utilizing modern ConvNet architectures and training techniques, state-of-the-art approaches~\citep{Nah_2017_CVPR,Tao_2017_ICCV,Gao_2019_CVPR,Yuan_2020_CVPR,Park_2020_ECCV_MTRNN,Chi_2021_CVPR} have been developed toward higher capacity and deblurring accuracy.
% Still, most methods tend to suffer from the blurry predictions due to the inherent limitation of PSNR-oriented solutions for ill-posed problems with large solution space~\citep{Ledig_2017_CVPR,Menon_2020_CVPR_pulse}.
Still, most methods tend to suffer from the blurry predictions due to the regression-to-mean behavior often witnessed in ill-posed problems with large solution space~\citep{Ledig_2017_CVPR,Menon_2020_CVPR_pulse}.

% the limitations of the previous perceptual losses
To overcome limitations of the conventional objectives, concepts of perceptual~\citep{Johnson2016Perceptual} and adversarial~\citep{Ledig_2017_CVPR,Nah_2017_CVPR,Kupyn_2018_CVPR} loss terms from high-level semantic tasks have been introduced to improve the visual quality of the deblurred results.
Nevertheless, such high-level losses may not serve as optimal goals for blur removal as low-level structural properties, e.g., blurriness, are not the primary features considered in their formulations.
As illustrated in Figure~\ref{fig:fig_kernel}, results from the previous deblurring methods are still blurry to a degree and the VGG and the adversarial losses are not sufficient to obtain perceptually pleasing and sharp images across different architectures~\citep{Tao_2018_CVPR,Gao_2019_CVPR,Kupyn_2019_ICCV}.

% Motivation: clean images are hard to reblur -> reblurring loss
While the deblurred images look less blurry compared with the original input, it is still possible to find nontrivial blur kernels with directional motion information.
% \sanghyun{While the deblurred images look less blurry compared with the original input, it is still possible to retrieve the blur kernel to some extent.}
% Despite the reduced blur strength in the deblurred images, the directional motion information remains observable in the blur kernels.
From the observation, we introduce the concept of \emph{reblurring} which amplifies the unremoved blur in the given image and reconstructs the original blur.
We note that our reblurring operation aims to recover the original motion trajectory in the blurry input, rather than to synthesize arbitrary, e.g., Gaussian, blurs.
Therefore, an ideally deblurred clean image is hard to reblur as no noticeable blur can be found to be amplified, making reblurring an ill-posed task.
In contrast, it is straightforward to predict the original shape of blur from insufficiently deblurred images as shown in Figure~\ref{fig:fig_kernel}.
% We propose to use the difference as the new optimization objective, \emph{reblurring loss} for the image deblurring problem.
We propose to use the difference between non-ideally deblurred image and the ideal sharp image in terms of reblurring feasibility 
as the new optimization objective, \emph{reblurring loss} for the image deblurring problem.

% Reblurring loss - supervised, self-supervised
The reblurring loss is realized by jointly training a pair of deblurring and reblurring modules.
% From a deblurred output, the reblurring module tries to make the reblurred image close to the original blurry image.
The reblurring module performs the inverse operation of deblurring, trying to reconstruct the original blurry image from a deblurred output.
% \sanghyun{Check here.} 
Using the property that the blurriness of a reblurred image depends on the sharpness quality of the deblurred result, we construct two types of loss functions.
% By using the property that the reblurred results should vary by the degree of blurriness of an image, we construct two types of loss functions.
During the joint training, \emph{supervised reblurring loss} compares the amplified blurs between the deblurred and the sharp image.
Complementing L1 intensity loss, the supervised reblurring loss guides the deblurring module to focus on and eliminate the remaining blur.
%While the training method is similar to the adversarial training of GANs~\citep{goodfellow2014generative}, the purposes and effects of the adversary are different.
% \sanghyun{Check here}
While our training strategy is similar to the adversarial training of GANs~\citep{goodfellow2014generative} in a sense that our deblurring and reblurring modules play the opposite roles, the purposes and effects of the adversary are different.
The reblurring loss concentrates on image blurriness regardless of image realism.
Furthermore, in contrast to the GAN discriminators that are not often used at test time, our reblurring module can be used to facilitate \emph{self-supervised reblurring loss}.
By making the deblurred image harder to reblur, the deblurring module can adaptively optimize itself without referring to the ground truth.
% \sanghyun{It is not clearly explained that how the `self-supervised' reblurring module is used (only supervised version is described above).}
% \sanghyun{Instead, our reblurring module can be used at test-time in a \emph{self-supervised manner}, by making deblurred image to resist against the reblurring process.
% By doing so, our reblurring module lets the deblurring result to be adaptively optimized to each input without requiring the ground-truth.}
% Furthermore, we apply \emph{self-supervised reblurring loss} at test-time so that the deblurred image would be impossible to be reblurred as a sufficiently sharp image would.
%The self-supervised reblurring loss lets the deblurring module to adaptively optimize to each input without ground truth.
% \sanghyun{While discriminator is not involved in the test-time...}
% Sanghyun: Move it to front
%In contrast, discriminators in GANs are not typically involved in the test time for adaptation.

% Generalization as loss function
Our reblurring loss functions provide additional optimization directives to the deblurring module and can be generally applied to any learning-based image deblurring methods.
With the proposed approach, we can derive sharper predictions from existing deblurring methods without modifying their architectures.
We summarize our contributions as follows:
\begin{compactitem}[$\bullet$]
    \item 
    Based on the observation that clean images are hard to reblur, we propose novel loss functions for image deblurring.
    Our reblurring loss reflects the preference for sharper images and contributes to visually pleasing deblurring results.

    \item 
    At test-time, the reblurring loss can be implemented without a ground-truth image.
    We perform test-time adaptive inference via self-supervised optimization with each input.
    
    \item
    Our method is generally applicable to any learning-based methods and jointly with other loss terms.
    Experiments show that the concept of reblurring loss consistently contributes to achieving state-of-the-art visual sharpness as well as LPIPS and NIQE across different model architectures.

\end{compactitem}

%% file: sections/figs/kernel.tex
\begin{figure}[t]
    \figspace
    \vspace{-1mm}
    \centering
    \newcommand{\ww}{0.197\linewidth}
    \newcommand{\awbegin}{29.2mm}
    \newcommand{\aw}{24.1mm}
    \addtocounter{subfigure}{0}

    \subfloat[\textbf{True Sharp} \label{fig:fig_kernel_sharp}]{\includegraphics[align=c, width=\ww]{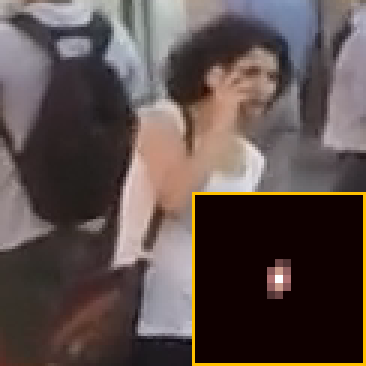}}
    % \hfill
    % \vrule
    \hfill
    \subfloat[SRN \label{fig:fig_kernel_srn}]{\includegraphics[align=c, width=\ww]{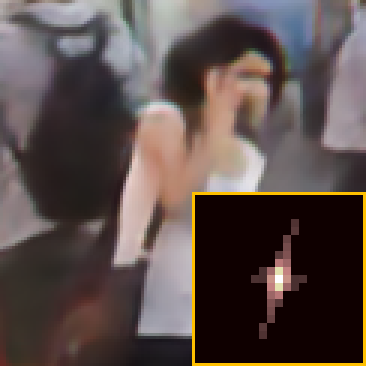}}
    \hfill
    \subfloat[SE-Sharing \label{fig:fig_kernel_se}]{\includegraphics[align=c, width=\ww]{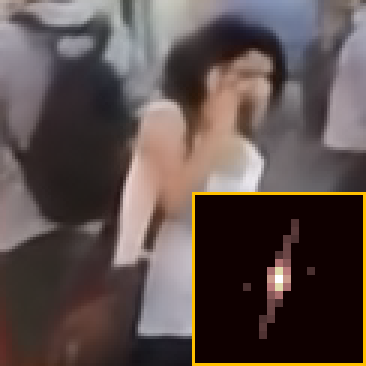}}
    \hfill
    \subfloat[DeblurGANv2 \label{fig:fig_kernel_deblurganv2}]{\includegraphics[align=c, width=\ww]{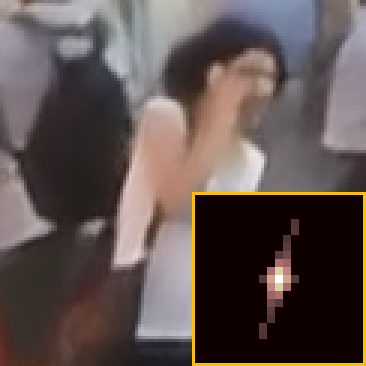}}
    \hfill
    \subfloat[\textbf{Ours} (Deblurred)]{\includegraphics[align=c, width=\ww]{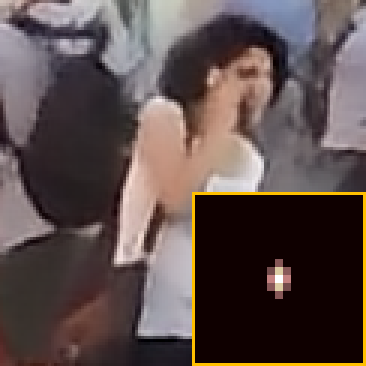}}
    \\
    % \vphantom{$\Downarrow$}
    % \hfill
    % \hphantom{$\Downarrow$} % (a)
    % \hfill
    % \vphantom{$\Downarrow$}
    % \hfill
    % $\Downarrow$            % (b)
    % \hfill
    % \vphantom{$\Downarrow$}
    % \hfill
    % $\Downarrow$            % (c)
    % \hfill
    % \vphantom{$\Downarrow$}
    % \hfill
    % $\Downarrow$            % (d)
    % \hfill
    % \vphantom{$\Downarrow$}
    % \hfill
    % $\Downarrow$            % (e)
    % \hfill
    % \vphantom{$\Downarrow$}
    % \\
    \hspace{\awbegin}
    $\Downarrow$
    \hspace{\aw}
    $\Downarrow$
    \hspace{\aw}
    $\Downarrow$
    \hspace{\aw}
    $\Downarrow$
    \\
    % \addtocounter{subfigure}{-5}
    \subfloat[\textbf{True Blur} \label{fig:fig_kernel_a}]{\includegraphics[align=c, width=\ww]{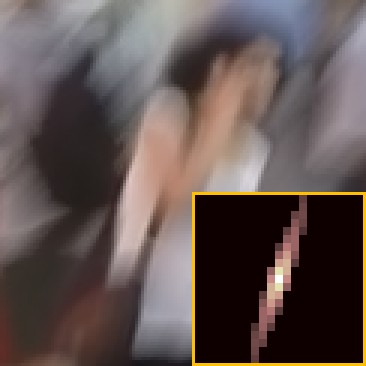}}
    % \hfill
    % \vrule
    \hfill
    \subfloat[Reblurred \label{fig:fig_kernel_b}]{\includegraphics[align=c, width=\ww]{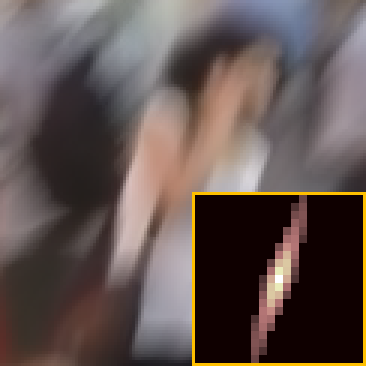}}
    \hfill
    \subfloat[Reblurred \label{fig:fig_kernel_c}]{\includegraphics[align=c, width=\ww]{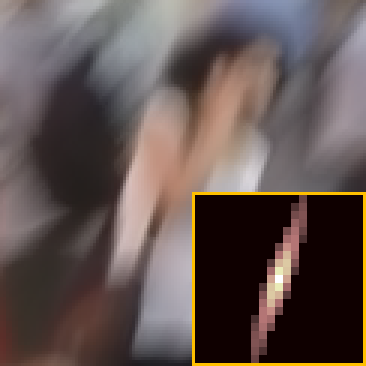}}
    \hfill
    \subfloat[Reblurred \label{fig:fig_kernel_d}]{\includegraphics[align=c, width=\ww]{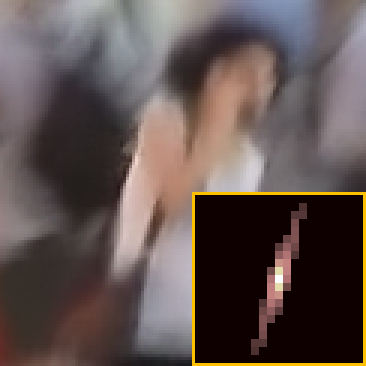}}
    \hfill
    \subfloat[\textbf{Ours} (Reblurred) \label{fig:fig_kernel_e}]{\includegraphics[align=c, width=\ww]{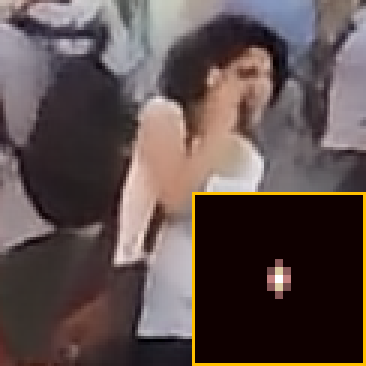}}
    \\
    \figcspace
    \caption{
        \textbf{Comparison of the deblurred images and their reblurred counterparts.}
        For each image, we visualize the remaining blur kernel~\citep{cho2009fast} at the center pixel visualized on the right bottom side.
        \textbf{Upper}: The kernels from the previous deblurring methods implicate the direction of the original blur.
        %
        %\textbf{Lower}: When the proposed reblurring module is applied, our result does not lose sharpness since the proposed method is designed to reconstruct outputs that are hard to be reblurred.
        \textbf{Lower}: When the proposed reblurring module is applied, our result does not lose sharpness as we reconstruct the output that is hard to be reblurred.
    }
    \label{fig:fig_kernel}
    \figspace
\end{figure}

%% file: sections/camera_ready/2_related_works.tex
\section{Related Works}
\label{sec:related_works}

\noindent \textbf{Image Deblurring.}
% Classic energy minimization -> prior design is important
Classical energy optimization framework is formulated by likelihood and prior terms.
Due to the ill-posedness of dynamic scene deblurring problem, prior terms have been essential in alleviating the optimization ambiguity, encoding the preference on the solutions.
Sophisticated prior terms were carefully designed with human knowledge on natural image statistics~\citep{levin2006blind,cho2009fast,hirsch2011fast,whyte2012non,sun2013edge,Xu_2013_CVPR,Kim_2013_ICCV,Kim_2014_CVPR,Pan_2016_CVPR}.
Recently in \citet{Li_2018_CVPR_discriminative_prior}, learned prior from a classifier discriminating blurry and clean images was also shown to be effective.
% In the recent work of \citet{Li_2018_CVPR_discriminative_prior}, learned prior, which is derived from a classifier discriminating blurry and clean images, was also shown to be effective.
Deep priors were also used for image deconvolution problems~\citep{Ren_2020_CVPR,Nan_2020_CVPR}.

% Deep learning methods -> focused on supervision term, architecture. no prior.
On the other hand, deep learning methods have benefited from learning on large-scale datasets.
The datasets consisting of realistic blur~\citep{Nah_2017_CVPR,Su_2017_CVPR,noroozi2017motion,Nah_2019_CVPR_Workshops_REDS,Gao_2019_CVPR,Jin_2019_CVPR,Shen_2019_ICCV_Human_aware} align the temporal center of the blurry and the sharp image pairs with high-speed cameras.
Learning from such temporally aligned data relieve the ill-posedness of deblurring compared with difficult energy optimization framework.
Thus, more attention has been paid to designing CNN architectures and datasets than designing loss terms.

% Various image deblurring CNN architectures
In the early work of \citet{schuler2015learning}, the alternating estimation of blur kernel and restored image~\citep{cho2009fast} was adopted in CNN architecture.
In \citet{Sun_2015_CVPR,Gong_2017_CVPR}, the spatially varying blur kernels are estimated by assuming locally linear blur followed by non-blind deconvolution.
Later, end-to-end learning without explicit kernel estimation became popular.
Motivated from the coarse-to-fine approach, multi-scale CNN was proposed~\citep{Nah_2017_CVPR} to expand the receptive field efficiently, followed by scale-recurrent architectures~\citep{Tao_2018_CVPR,Gao_2019_CVPR}.
On the other hand, \citet{Zhang_2019_CVPR_DMPHN,Suin_2020_CVPR_SAPHN} sequentially stacked network modules.
Recently, \citet{Park_2020_ECCV_MTRNN} proposed a multi-temporal model that deblurs an image recursively.
To handle spatially varying blur kernels efficiently, spatially non-uniform operations were embedded in neural networks~\citep{Zhang_2018_CVPR_svrn,Yuan_2020_CVPR}.

\noindent \textbf{Perceptual Image Restoration.}
% conventional L1, L2 optimization
Often, L1 or L2 losses are used at training to achieve higher PSNR.
% Conventional image restoration methods mainly optimize L1 or L2 loss to achieve higher PSNR.
However, such approaches suffer from blurry and over-smoothed outputs~\citep{Johnson2016Perceptual, Zhang_2018_CVPR_unreasonable, Menon_2020_CVPR_pulse} as the learned models predict an average of all possible solutions under the ill-posedness~\citep{Ledig_2017_CVPR}.
% The primary reason is that the learned models predict an average of all possible solutions under the ill-posedness~\citep{Ledig_2017_CVPR}.
To deal with the issue, several studies utilize deep features of the pretrained VGG~\citep{simonyan2014very} and other networks that are more related to human perception~\citep{Johnson2016Perceptual, Zhang_2018_CVPR_unreasonable} and with analysis on frequency space~\citep{tariq2020deep,czolbe2020loss}.
% Then, the following methods can produce perceptually better results by minimizing the distance of output and ground-truth images in the feature domain.
Recent methods introduce adversarial training~\citep{goodfellow2014generative} so that outputs of the restoration models be indistinguishable from real samples~\citep{Nah_2017_CVPR,Nimisha_2017_ICCV,Ledig_2017_CVPR,Kupyn_2018_CVPR,Kupyn_2019_ICCV}.
Also, there were attempts to exploit statistical properties of images and features with contextual loss~\citep{mechrez2018maintaining} and projected distribution loss~\citep{delbracio2020projected}.

% there were attemps to use image and feature statistics such as contextual loss~\citep{mechrez2018maintaining}, projected distribution loss~\cite{delbracio2020projected}

Nevertheless, an inherent limitation of existing perceptual objectives is that they are not task-specialized for image restoration.
For example, the VGG features are learned for high-level visual recognition while the adversarial loss only contributes to reconstructing realistic images without considering the existence of motion blur.
Therefore, blindly optimizing those terms may not yield an optimal solution in terms of image deblurring.
In practice, we observed that those objectives still tend to leave blur footprints unremoved, making it possible to estimate the original blur.
Our reblurring loss is explicitly designed to improve the perceptual sharpness of deblurred images by reducing remaining blurriness and thus more suitable for deblurring, acting as a learned prior.

% perceptual image restoration works
% \citep{mechrez2018maintaining}

% \citep{delbracio2020projected}

% \citep{tariq2020deep}

% \citep{czolbe2020loss}

\noindent \textbf{Image Blurring.}
As an image could be blurred in various directions and strength, image blurring is another ill-posed problem without additional information.
% Thus, intrinsic~\citep{bae2007defocus} or extrinsic~\citep{chen2018reblur2deblur,Brooks_2019_CVPR_synthesize_motion} information is often incorporated.
Thus, intrinsic or extrinsic information is often incorporated.
With a non-ideally sharp image, \citet{bae2007defocus} detected the small local blur kernel in the image to magnify the defocus blur for bokeh effect.
On the other hand, \citet{chen2018reblur2deblur} estimated the kernel by computing the optical flow from the neighboring video frames.
Similarly, \citet{Brooks_2019_CVPR_synthesize_motion} used multiple video frames to synthesize blur.
% Without such blur or motion cue, there could be infinitely many types of plausible blur applicable to an image.
% Thus, \citet{Zhang_2020_CVPR_BGAN} used a generative model to synthesize many blurry images.
Without such external information, \citet{Zhang_2020_CVPR_BGAN} used a generative model to synthesize many blurry images.
In contrast, \citet{Bahat_2017_ICCV} deliberately blurred an already blurry image in many ways to find the local blur kernel.
Our image reblurring concept is similar to \citet{bae2007defocus} in the sense that intrinsic cue in an image is used to amplify blur.
Nonetheless, our main goal is to use reblurring to provide a guide to deblurring model so that such blur cues would be better removed.

%% file: sections/camera_ready/3_methods.tex
\section{Proposed Method}
\label{sec:method}

\input{sections/figs/reb_deb}

In this section, we describe a detailed concept of image reblurring and how the operation can be learned.
Then, we demonstrate that the operation can support the deblurring module to reconstruct perceptually favorable and sharp images.
We also propose a self-supervised test-time optimization strategy by using the learned reblurring module.
%The proposed reblurring loss can support the deblurring modules to reconstruct perceptually favorable and sharp outputs.
% \sanghyun{Do we need the next sentence here?}
% At training and testing stages, we formulate the reblurring loss in supervised and self-supervised manner.
For simplicity, we refer to the blurry, the deblurred, and the sharp image as $B$, $L$, and $S$, respectively.

\input{sections/tables/psnr_deb_reb}

\subsection{Clean Images are Hard to Reblur}
\label{sec:cihr}

% motivation: a clean image is hard to reblur
As shown in Figure~\ref{fig:fig_kernel}, outputs from the existing deblurring methods still contain undesired motion trajectories that are not completely removed from the input.
Ideally, a well-deblurred image should not contain any motion cues, making reblurring ill-posed.
We first validate our motivation by buliding a reblurring module $\mathcal{M}_{\text{R}}$ which amplifies the remaining blur from $L$.
% To validate our motivation that clean images are hard to reblur, we first build a reblurring module $\mathcal{M}_{\text{R}}$ which amplifies the remaining blur from $L$.
$\mathcal{M}_{\text{R}}$ is trained with the following blur reconstruction loss $\mathcal{L}_{\text{Blur}}$ so that it would learn the inverse operation of deblurring.
\begin{equation}
    \mathcal{L}_\text{Blur} = \norm{\mathcal{M}_{\text{R}}(L) - B}.
    \label{eq:eq_loss_br}
\end{equation}
We apply $\mathcal{M}_{\text{R}}$ to the deblurred images from deblurring modules of varying capacities.
Table~\ref{tab:deblur_reblur_psnr} shows that the higher the deblurring PSNR, the lower the reblurring PSNR becomes when both modules are trained with conventional L1 loss, independently from each other.
% \sanghyun{Check here.} 
It demonstrates that the better deblurred images are harder to reblur, consistent to our motivation.
% It demonstrates the better deblurred images are harder to reblur, justifying our motivation.

In contrast to the non-ideally deblurred images, $\mathcal{M}_{\text{R}}$ is not able to generate a motion blur from a sharp image $S$ as no motion information is found.
For a high-quality clean image, $\mathcal{M}_{\text{R}}$ should preserve the sharpness.
However, optimizing the blur reconstruction loss $\mathcal{L}_{Blur}$ with $S$ may fall into learning the pixel average of all blur trajectories in the training dataset, i.e. Gaussian blur.
In such a case, $\mathcal{M}_{R}$ will apply the single uniform blur on every image without considering the scene information.
To let the blur domain of $\mathcal{M}_{R}$ confined to the motion-incurred blur, we use sharp images to penalize such undesired operations.
Specifically, we introduce a network-generated sharp image $\hat{S}$ obtained by feeding a real sharp image $S$ to the deblurring module $\mathcal{M}_{\text{D}}$ as $\hat{S}=\mathcal{M}_{\text{D}}(S)$.
We define sharpness preservation loss $\mathcal{L}_{\text{Sharp}}$ as follows:
\begin{equation}
    \mathcal{L}_{\text{Sharp}} = \norm{\mathcal{M}_{\text{R}}(\hat{S}) - \hat{S}}.
    \label{eq:eq_loss_sp}
\end{equation}
We use the pseudo-sharp image $\hat{S}$ instead of a real image $S$ to make our reblurring module focus on image sharpness and blurriness rather than the realism.
While $\hat{S}$ differ from $L$ only by the sharpness, $S$ also differ by the realism which can be easily detected by neural networks~\citep{Wang_2020_CVPR_easy}.

Combining both terms together, we train the reblurring module $\mathcal{M}_\text{R}$ by optimizing the joint loss $\mathcal{L}_\text{R}$:
\begin{equation}
    \mathcal{L}_{\text{R}} = \mathcal{L}_{\text{Blur}} + \mathcal{L}_{\text{Sharp}}.
    \label{eq:eq_loss_R}
\end{equation}
As zero-magnitude blur should remain unaltered from $\mathcal{M}_{\text{R}}$, the sharpness preservation loss can be considered a special case of the blur reconstruction loss.
Figure~\ref{fig:fig_reblur} illustrates the way our reblurring module is trained from $\mathcal{L}_{\text{R}}$.

\subsection{Supervision from Reblurring Loss}
\label{sec:reblur_loss_supervised}

\input{sections/figs/concept}

% supervised reblurring loss
The blurriness of images can be more easily witnessed by amplifying the blur.
Thus, we propose a new optimization objective by processing the deblurred and the sharp image with the jointly trained reblurring model $\mathcal{M}_{\text{R}}$.
To suppress the remaining blur in the output image $L=\mathcal{M}_\text{D}(B)$ from the deblurring module $\mathcal{M}_\text{D}$, the \emph{supervised reblurring loss} $\mathcal{L}_\text{Reblur}$ for image deblurring is defined as
\begin{equation}
    \mathcal{L}_{\text{Reblur}} = \norm{\mathcal{M}_{\text{R}}(L) - \mathcal{M}_{\text{R}}(S)}.
    \label{eq:eq_loss_reblur_sup}
\end{equation}
%
% why S^ was not used
Unlike the sharpness preservation term in \eqref{eq:eq_loss_sp}, we do not use the pseudo-sharp image $\hat{S}$ in our reblurring loss, $\mathcal{L}_{\text{Reblur}}$.
As the quality of the pseudo-sharp image $\hat{S}$ depends on the state of deblurring module $\mathcal{M}_{\text{D}}$, using $\hat{S}$ may make training unstable and difficult to optimize, especially at the early stage.
Thus we use a real sharp image $S$ to stabilize the training.
Nevertheless, as $\mathcal{M}_{\text{R}}$ is trained to focus on the sharpness from \eqref{eq:eq_loss_R}, so does the reblurring loss, $\mathcal{L}_{\text{Reblur}}$.

% total loss for deblurring
Using our reblurring loss in \eqref{eq:eq_loss_reblur_sup}, the deblurring module $\mathcal{M}_\text{D}$ is trained to minimize the following objective $\mathcal{L}_\text{D}$:
\begin{equation}
    \mathcal{L}_{\text{D}} = \mathcal{L}_{1} + \lambda \mathcal{L}_{\text{Reblur}},
    \label{eq:eq_loss_deblur}
\end{equation}
where $\mathcal{L}_1$ is a conventional L1 loss, and the hyperparameter $\lambda$ is empirically set to 1.
Figure~\ref{fig:fig_deblur} illustrates how the deblurring model is trained with the guide from $\mathcal{M}_{\text{R}}$.

% differences from adversarial loss
At each training iterations, we alternately optimize two modules $\mathcal{M}_{\text{D}}$ and $\mathcal{M}_{\text{R}}$ by $\mathcal{L}_{\text{D}}$ and $\mathcal{L}_{\text{R}}$, respectively.
While such a strategy may look similar to the adversarial training scheme, the optimization objectives are different.
As the neural networks are well known to easily discriminate real and fake images~\citep{Wang_2020_CVPR_easy}, the realism could be as a more obvious feature than image blurriness.
Thus, adversarial loss may overlook image blurriness as $L$ and $S$ can already be discriminated by the difference in realism.
On the other hand, our reblurring loss is explicitly designed to prefer sharp images regardless of realism as we use $\hat{S}$ instead of $S$ in $\mathcal{L}_{\text{Sharp}}$ to train $\mathcal{M}_{\text{R}}$.
Figure~\ref{fig:fig_concept} conceptually compares the actual role of the reblurring loss $\mathcal{L}_{\text{Reblur}}$ and the adversarial loss $\mathcal{L}_{\text{Adv}}$.

\subsection{Test-time Adaptation by Self-Supervision}
\label{sec:tta}

% necessity for self-supervised optimization
After the training is over, the models learned from supervised loss terms have fixed weights at test time.
% After we optimize the deblurring network with \eqref{eq:eq_loss_deblur}, the model has fixed weights during the test time.
When a new example that deviates from the distribution of training data is given, the supervised methods may lack ability to generalize.
Our reblurring module, however, can further provide self-supervised guide so that the model could be further optimized for each image at test time.
While the supervised reblurring loss $\mathcal{L}_{\text{Reblur}}$ finds the blurriness of $L$ by comparison with the ground truth, $\mathcal{M}_{\text{R}}$ can also inspect the blurriness of an image without reference.

% the self-supervised test-time adaptation
As $\mathcal{M}_{\text{R}}$ is trained to magnify the blur in $L$, imperfectly deblurred image would be blurred.
Thus, the difference between $\mathcal{M}_{\text{R}}(L)$ and $L$ can serve as a feedback without having to reference $S$.
Furthermore, due to the sharpness preservation loss $\mathcal{L}_{\text{Sharp}}$, a sufficiently sharp image would have little difference when reblurred.
Based on the property, we construct the \emph{self-supervised reblurring loss} that serves as a prior term embedding the preference on sharp images as
\begin{equation}
    \mathcal{L}_{\text{Reblur}}^{\text{self}} = \norm{ \mathcal{M}_{\text{R}}(L) - L_{\ast} },
    \label{eq:eq_loss_reblur_self}
\end{equation}
where $L_{\ast}$ denotes the image with the same value as $L$ but the gradient does not backpropagate in the optimization process.
We minimize $\mathcal{L}_{\text{Reblur}}^{\text{self}}$ for each test data to obtain the sharper image.
Allowing gradient to flow through $L_{\ast}$ can let $L$ to fall into undesired local minima where both the $L$ and $\mathcal{M}_{\text{R}}(L)$ are blurry.
We iteratively optimize the weights of $\mathcal{M}_{\text{D}}$ with fixed $\mathcal{M}_{\text{R}}$.
As $\mathcal{L}_{\text{Reblur}}^{\text{self}}$ only considers the sharpness of an image, we keep the color consistency by matching the color histogram between the test-time adapted image and the initially deblurred image.
For the detailed optimization process of test-time adaptation strategy, please refer to the Appendix Algorithm~\ref{alg:tta}.
The effect of test-time adaptation is conceptually visualized in Figure~\ref{fig:fig_tta}.
More iterations make the deblurred image sharper.
We note that our loss functions and the test-time adaptation are applicable to general learning-based approaches.

%% file: sections/figs/reb_deb.tex
\begin{figure*}[t!]
    \vspace{-4mm}
    \subfloat[Reblurring module training process.
    % $\mathcal{M}_{\text{R}}$ tries to reconstruct the original blur $B$ from an imperfect deblurred image $L$, while preserving the sharpness of a pseudo-sharp image $\hat{S}$.
    \label{fig:fig_reblur}]{\includegraphics[width=.45\linewidth]{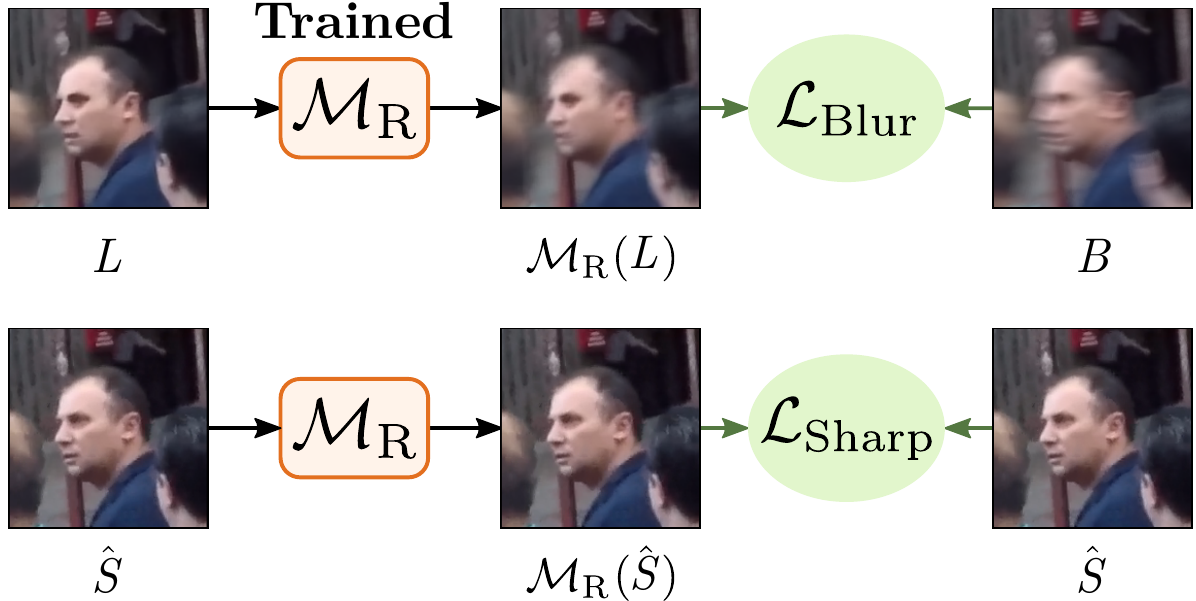}}
    \hfill
    \vrule depth -0.01cm
    \hfill
    \subfloat[Image deblurring with reblurring loss \label{fig:fig_deblur}]{\includegraphics[width=.52165\linewidth]{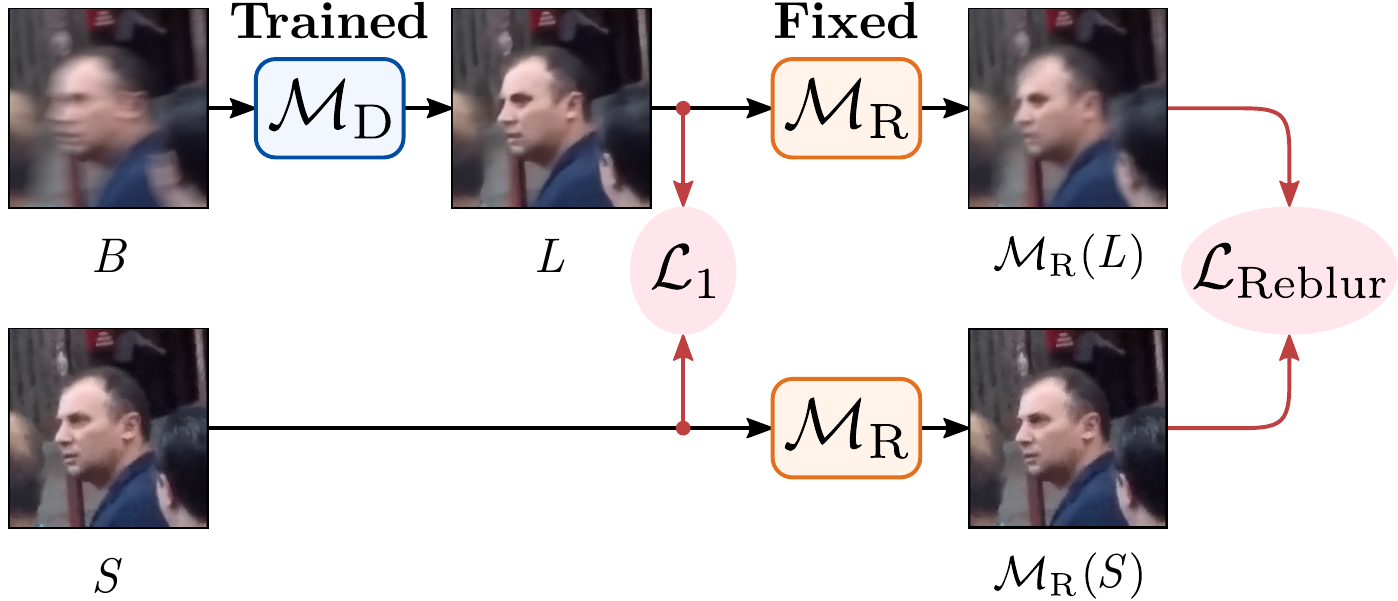}}
    \figcspace
    \caption{
        \textbf{Overviews of the proposed reblurring and deblurring framework.}
        Reblurring module $\mathcal{M}_{\text{R}}$ tries to reconstruct blurry image $B$ from a deblurred image $L$ while preserving the sharpness of a pseudo-sharp image $\hat{S}=\mathcal{M}_{\text{D}}(S)$.
        Meanwhile, the deblurring module $\mathcal{M}_{\text{D}}$ tries to make $L$ sharper by comparing the amplified blur from $L$ and the sharp image $S$.
    }
    \label{fig:fig_reb_deb}
    % \figspace
    \vspace{-2mm}
\end{figure*}

%% file: sections/tables/psnr_deb_reb.tex
\begin{table}[t!]
    \centering
    \footnotesize
    % \begin{tabularx}{\linewidth}{l | >{\centering\arraybackslash}X | >{\centering\arraybackslash}X | >{\centering\arraybackslash}X | >{\centering\arraybackslash}X}
    \begin{tabularx}{\linewidth}{l  >{\centering\arraybackslash}X  >{\centering\arraybackslash}X  >{\centering\arraybackslash}X  >{\centering\arraybackslash}X}
        \toprule
        % $\mathcal{M}_{\text{D}}$ \#ResBlocks & 4 & 8 & 16 & 32\\
        % \#ResBlocks in $\mathcal{M}_{\text{D}}$ & 4 & 8 & 16 & 32\\
        \#ResBlocks in deblurring module $\mathcal{M}_{\text{D}}$ & 4 & 8 & 16 & 32\\
        \midrule
        Deblur PSNR wrt sharp GT & 28.17 & 29.67 & 30.78 & 31.48\\
        Reblur PSNR wrt blur GT & 34.29 & 32.66 & 31.90 & 31.48\\
        \bottomrule
    \end{tabularx}
    \tabcspace
    \caption{
        \textbf{Deblurring and reblurring PSNR (dB) by deblurring model capacity.}
        Both tasks are trained independently with L1 loss on the GOPRO dataset.
        % We note that \#ResBlocks varies in the deblur network only.
        The number of ResBlocks in $\mathcal{M}_{\text{R}}$ is 2.
    }
    \label{tab:deblur_reblur_psnr}
    \tabspace
\end{table}

%% file: sections/figs/concept.tex
\begin{figure}[t]
\vspace{-4mm}
\renewcommand{\wp}{0.49}
    \begin{center}

        \subfloat[Deblurring and Reblurring operations\label{fig:fig_concept_deblur_reblur}]{\includegraphics[width=\wp\linewidth]{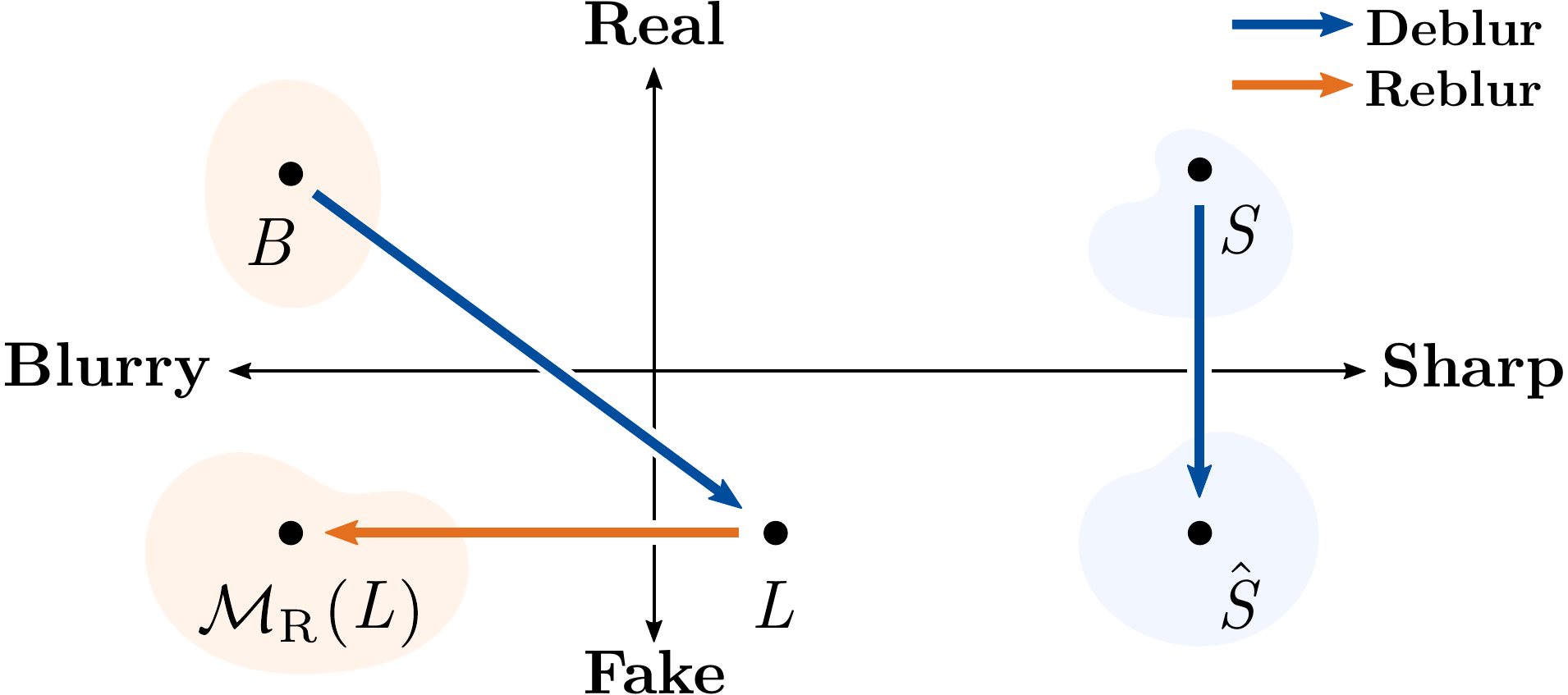}}
        \hfill
        \subfloat[Reblurring and Adversarial losses\label{fig:fig_concept_loss_comp}]{
        \includegraphics[width=\wp\linewidth]{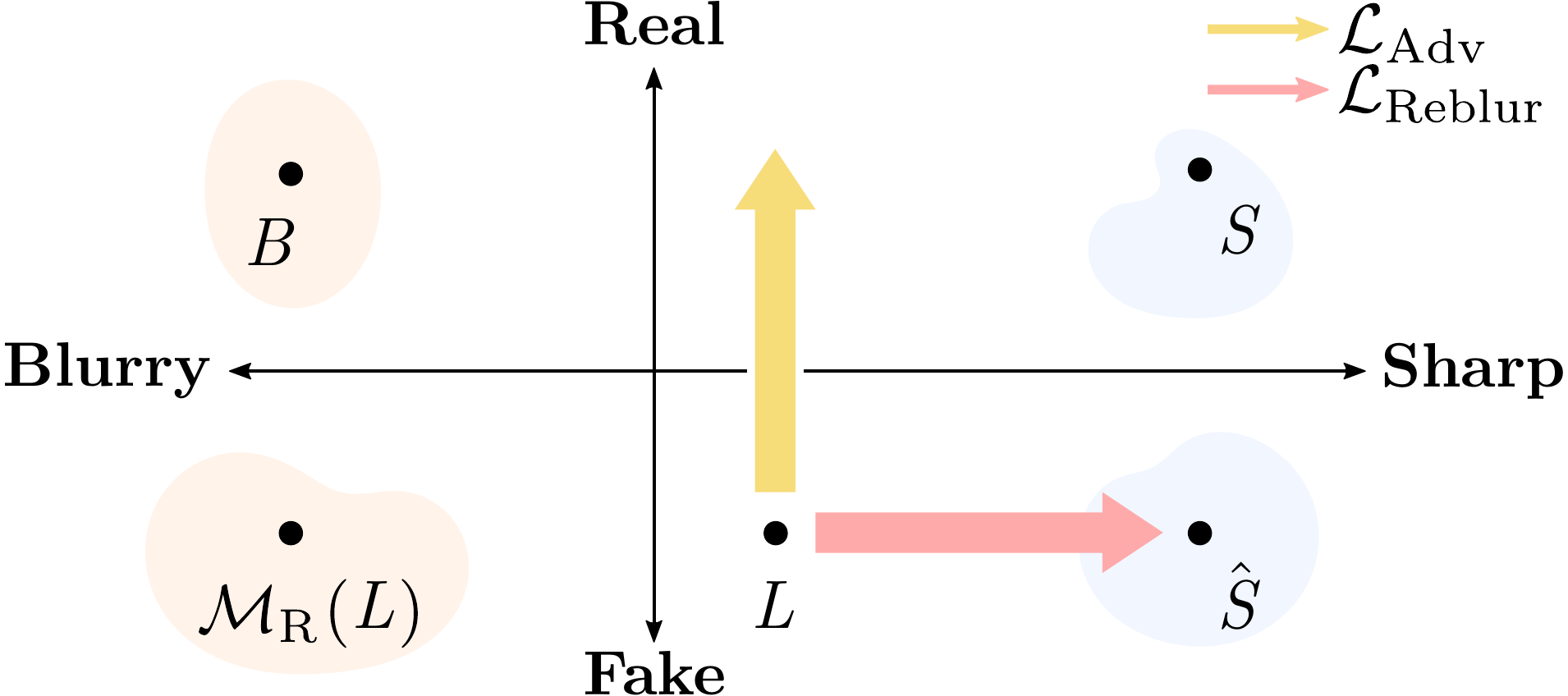}}
        
    \end{center}
    \figcspace
    \caption{
        \textbf{Image deblurring and reblurring illustrated from the perspective of sharpness and realism.}
        Training our modules with $\mathcal{L}_{\text{Reblur}}$ improves image sharpness without considering the image realism.
        The image realism can be optionally handled by adversarial loss $\mathcal{L}_\text{Adv}$.
    }
    \label{fig:fig_concept}
    \figspace
\end{figure}

% \begin{figure}[t]
%     \begin{minipage}[t]{0.66 \linewidth}
%         \centering
%         % \includegraphics[width=\linewidth]{figs/fig_concept/fig_concept_v16.pdf}
%         \includegraphics[width=.49\linewidth]{figs/fig_concept/fig_concept_model_v1.pdf}
%         \includegraphics[width=.49\linewidth]{figs/fig_concept/fig_concept_loss_v1.pdf}
%         \figcspace
%         \caption{
%             \textbf{Deblurring and reblurring operations illustrated with the role of different loss functions in terms of sharpness and realism.}
%             % \textbf{Image deblurring and reblurring illustrated from the perspective of sharpness and realism.}
%             % Training our modules with $\mathcal{L}_{\text{Reblur}}$ improves image sharpness without considering the image realism.
%             % The image realism can be optionally handled by adversarial loss $\mathcal{L}_\text{Adv}$.
%         }
%         \label{fig:fig_concept}
%     \end{minipage}
%     \hfill
%     \begin{minipage}[t]{0.33 \linewidth}
%         \centering
%         \includegraphics[width=\linewidth]{figs/fig_tta/fig_tta_v3.pdf}
%         % \includegraphics[width=\linewidth]{figs/fig_tta/fig_tta_v3.pdf}
%         \figcspace
%         \caption{
%             \textbf{The proposed self-supervised test-time adaptation.}
%             We repetitively find the latent image that reblurs to the current deblurred image.
%         }
%         \label{fig:fig_tta}
%     \end{minipage}
%     \figspace
% \end{figure}

%% file: sections/camera_ready/4_experiments.tex
\section{Experiments}
\label{sec:experiments}
% \sanghyun{Check here.}
We verify the effectiveness of our reblurring loss and the generalization by applying it to multiple deblurring model architectures.
% We demonstrate the effectiveness of our reblurring loss by applying it to multiple deblurring model architectures.
We show the experimental results with a baseline residual U-Net and state-of-the-art image deblurring models, the sRGB version SRN~\citep{Tao_2018_CVPR} and DHN, our modified version of DMPHN~\citep{Zhang_2019_CVPR_DMPHN}.
For the reblurring module, we use simple residual networks with 1 or 2 ResBlock(s) with $5 \times 5$ convolution kernels.
The training and evaluation were done with the widely used GOPRO~\citep{Nah_2017_CVPR} and REDS~\citep{Nah_2019_CVPR_Workshops_REDS} datasets.
The GOPRO dataset consists of 2103 training and 1111 test images with various dynamic motion blur.
Similarly, the REDS dataset has 24000 training and 3000 validation data publicly available.
For each dataset, every experiment was done in the same training environment.
We mainly compare LPIPS~\citep{Zhang_2018_CVPR_unreasonable} and NIQE~\citep{mittal2012making} perceptual metrics as our goal is to make images sharper.
For more implementation details, please refer to the Appendix.

\input{sections/tables/tta_gopro_reds}

\subsection{Effect of Supervised Reblurring Loss}
\label{sec:experiment_reblur_supervised}

We implement the reblurring loss in varying degrees of emphasis on sharpness by controlling the reblurring module capacity.
For a more balanced quality between PSNR and perceptual sharpness, we use 1 ResBlock for $\mathcal{M}_{\text{R}}$.
To put more weight on the perceptual quality, we allocate a larger capacity on $\mathcal{M}_{\text{R}}$ by using 2 ResBlocks.
For notation simplicity, we denote the reblurring loss with $k$ ResBlock(s) in the reblurring module as $\mathcal{L}_{\text{Reblur, n}k}$.

Table~\ref{tab:tta_comparison} shows how the deblurring performance varies by the use of reblurring loss functions.
With $\mathcal{L}_{\text{Reblur, n1}}$, LPIPS and NIQE improves to a moderate degree while PSNR and SSIM metrics remain at a similar level.
Meanwhile, $\mathcal{L}_{\text{Reblur, n2}}$ more aggressively optimizes the perceptual metrics.
This is analogous to the perception-distortion trade-off witnessed in the image restoration literature~\citep{blau2018perception,Blau_2018_ECCV_Workshops}.
The perceptual metric improvements are consistently witnessed with various architectures on both the GOPRO and the REDS dataset.

\input{sections/figs/loss_compare}

\subsection{Effect of Sharpness Preservation Loss}
\label{sec:sharpness_preservation}

In training $\mathcal{M}_{\text{R}}$, we used both the blur reconstruction loss $\mathcal{L}_\text{Blur}$ and the sharpness preservation loss $\mathcal{L}_\text{Sharp}$.
The latter term $\mathcal{L}_\text{Sharp}$ plays an essential role in letting $\mathcal{M}_{\text{R}}$ concentrate only on the motion-driven blur in the given image and keep sharp image remain sharp.
Table~\ref{tab:gopro_sharpness_preservation} presents the performance gains from using $\mathcal{L}_\text{Sharp}$ jointly with $\mathcal{L}_\text{Blur}$ in terms of the perceptual quality.

Also, the use of the pseudo-sharp image $\hat{S}$ is justified by comparing it with the use case of real sharp image $S$.
% Table~\ref{tab:gopro_sharpness_preservation} also justifies the effectiveness of the pseudo-sharp image $\hat{S}$ in sharpness preservation.
We found the using $S$ for $\mathcal{L}_\text{Sharp}$ with $\mathcal{L}_\text{Blur}$ makes the training less stable than using $\hat{S}$.
Using the pseudo-sharp image confines the input data distribution of $\mathcal{M}_{\text{R}}$ to the domain of $\mathcal{M}_{\text{D}}$ outputs.
For neural networks, it is very easy to discriminate real and fake images~\citep{Wang_2020_CVPR_easy}.
% By using a fake image $\hat{S}$ instead of a real image $S$, we let $\mathcal{M}_{\text{R}}$ focus on the sharpness of an image and avoid being distracted by the difference in image realism which is more obvious to it.
By using a fake image $\hat{S}$ instead of a real image $S$, we let $\mathcal{M}_{\text{R}}$ focus on the sharpness of an image and avoid being distracted by a more obvious difference between real and fake images.
% Otherwise, $\mathcal{M}_{\text{R}}$ could discriminate $L$ and $S$ as fake and real images and ignore the blurriness difference.
% In contrast to $S$ that differ from the deblurred image $L$ by the realness, $\hat{S}$ can avoid being $\mathcal{M}_{\text{R}}$ distracted by such an unintentional difference, focusing on image sharpness.
% While the real sharp data $S$ differ from the deblurred image $L$ in terms of realness, the pseudo-sharp image $\hat{S}$ only differs by the sharpness.
% Thus the reblurring module can focus on the image sharpness without being distracted by other unintended properties.
Furthermore, it leads the two loss terms $\mathcal{L}_{\text{Blur}}$ and $\mathcal{L}_{\text{Sharp}}$ to reside under the same objective, amplifying any noticeable blur and keeping sharpness when motion blur is in zero-magnitude.

\subsection{Comparison with Other Perceptual Losses}
\label{sec:comp_reblur_adv}
The reblurring loss provides a conceptually different learning objectives from the adversarial and the perceptual losses and is designed to focus on the motion blur.
Table~\ref{tab:comp_reblur_adv} compares the effect of $\mathcal{L}_{\text{Reblur}}$ with adversarial loss $\mathcal{L}_{\text{Adv}}$, and the VGG perceptual loss~\citep{Johnson2016Perceptual} by applying them to SRN~\citep{Tao_2018_CVPR} on GOPRO dataset.
While our method provides quantitatively better perceptual scores, the different perceptual losses are oriented to different goals.
They do not necessarily compete or conflict with each other and can be jointly applied at training to catch the perceptual quality in varying aspects.
In Figure~\ref{fig:fig_reblur_loss_gopro}, the effect of the reblurring loss is visually compared with the previous perceptual loss functions.

\input{sections/tables/percep_loss_comp}

\subsection{Effect of Test-time Adaptation}
\label{sec:tta_effect}

\begin{figure}[t]
    \begin{minipage}[t]{0.51 \linewidth}
        \centering
        \includegraphics[width=\linewidth]{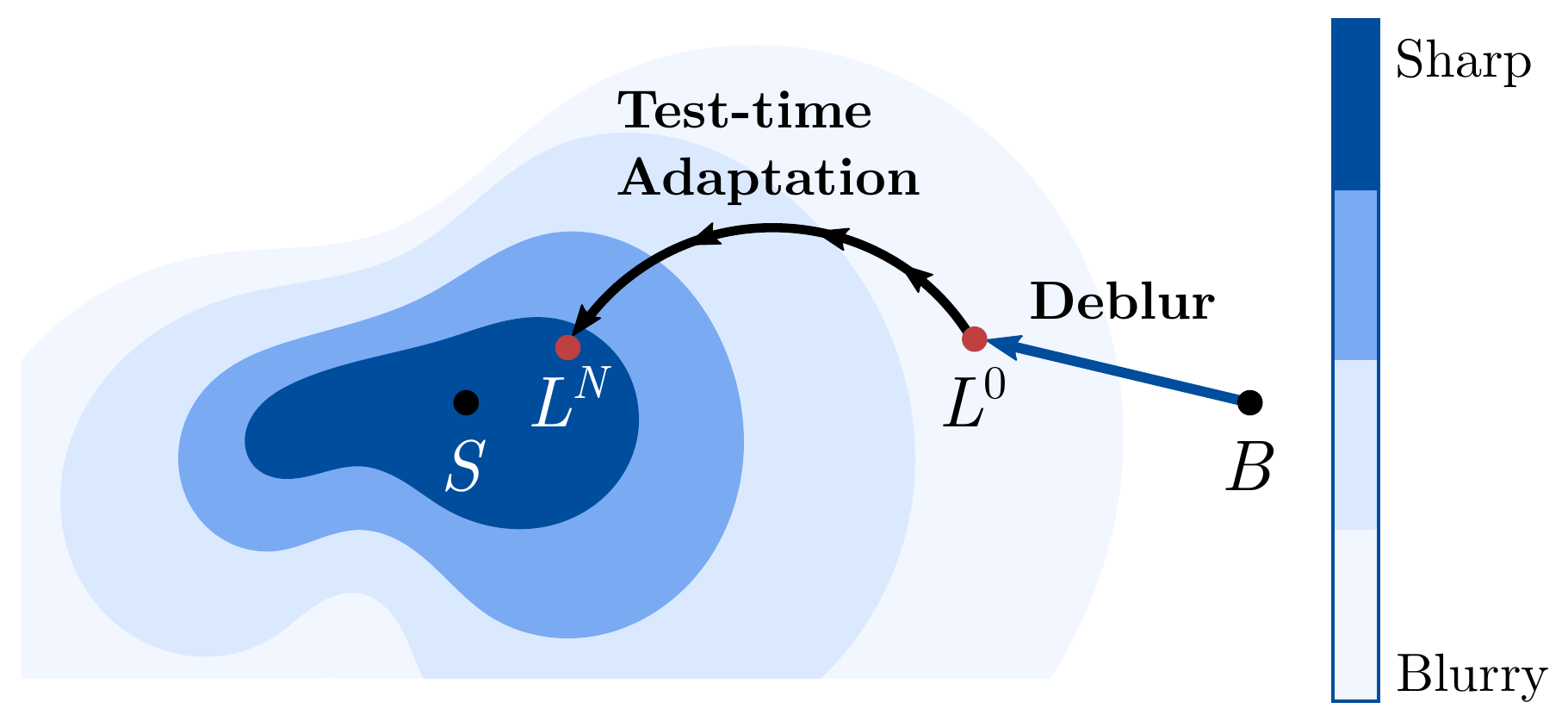}
        \\
        \vspace{-1.5mm}
        %\figcspace
        \caption{
        \textbf{The proposed self-supervised test-time adaptation.}
            The iterative optimization improves the image sharpness by finding an image that reblurs to the current deblurred image.
            % We repetitively find the latent image that reblurs to the current deblurred image.
        }
        \label{fig:fig_tta}
    \end{minipage}
    \hfill
    \begin{minipage}[t]{0.47 \linewidth}
        \centering
        \includegraphics[width=\linewidth]{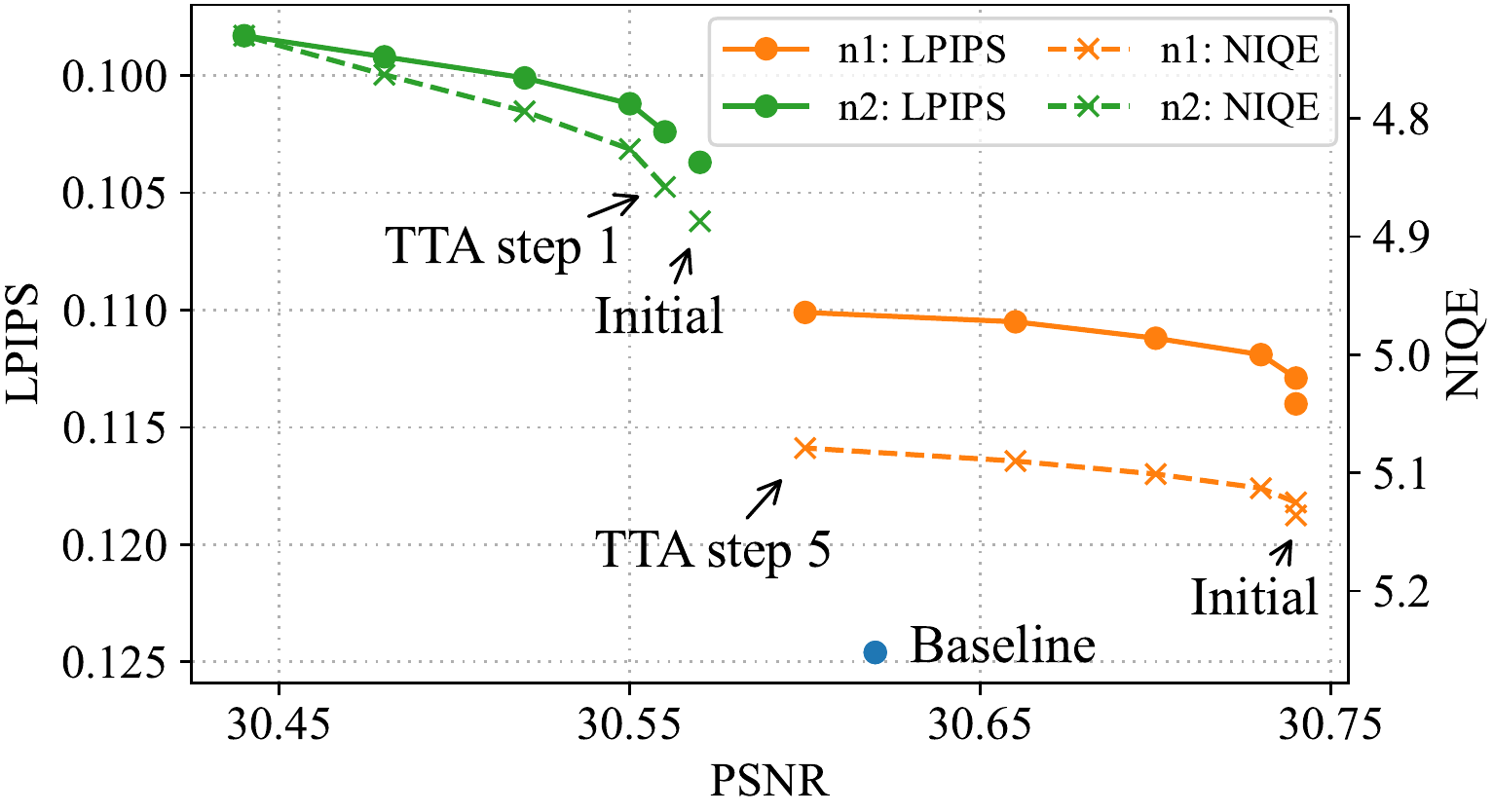}
        \\
        %\figcspace
        \vspace{-1.5mm}
        \caption{
            \textbf{Test-time adaption (SRN) on GOPRO dataset.}
            % Our self-supervised objective improves trade-off between the perceptual image quality (LPIPS, NIQE) and PSNR compared with the baseline.
            Reblurring loss improves the trade-off between the perception (LPIPS, NIQE) and PSNR compared with the baseline.
        }
        \label{fig:fig_psnr_lpips}
    \end{minipage}
    \vspace{-2mm}
    %\figspace
\end{figure}

We conduct test-time adaptation with the proposed self-supervised reblurring loss, $\mathcal{L}_{\text{Reblur}}^{\text{self}}$ to make the deblurred image even sharper.
Figure~\ref{fig:fig_psnr_lpips} shows the test-time-adapted result with SRN.
Compared with the baseline trained with L1 loss, our results exhibit improved trade-off relations between PSNR and the perceptual metrics, LPIPS and NIQE.
Table~\ref{tab:tta_comparison} provides detailed quantitative test-time adaptation results on GOPRO and REDS dataset, respectively with various deblurring module architectures.
The effect of test-time adaptation is visually shown in Figure~\ref{fig:fig_tta_reds}.

\subsection{Comparison with State-of-The-Art Methods}

We have improved the perceptual quality of the deblurred images by training several different model architectures.
We compare the perceptual quality with the other state-of-the-art methods in Figure~\ref{fig:fig_gopro}.
Especially, DeblurGAN-v2 was trained with the VGG loss and the adversarial loss.
Our results achieve visually sharper texture from the reblurring loss and test-time adaptation.

% Our models trained with reblurring loss shows visually sharper results.

% We compare our method with the state-of-the-art single image deblurring methods in Figure~\ref{fig:fig_gopro}.
% Our results from test-time adaptation recover the shape of fast-moving cars and the walking pedestrians' feet.
% While other methods exhibit higher PSNR and SSIM, our test-time adaptation can make the deblurred images visually much sharper.

\subsection{Real World Image Deblurring}

While our method uses synthetic datasets~\citep{Nah_2017_CVPR,Nah_2019_CVPR_Workshops_REDS} for training, the trained models generalize to real blurry images.
In Figure~\ref{fig:fig_real}, we show deblurred results from \citet{Lai_2016_CVPR} dataset with DHN model.
Compared with the baseline $\mathcal{L}_{1}$ loss, our reblurring loss $\mathcal{L}_{\text{Reblur, n2}}$ provides an improved deblurring quality.
As the real test image could deviate from the training data distribution, a single forward inference may not produce optimal results.
With the self-supervised test-time adaptation, our deblurred images reveal sharper and detailed textures.

\input{sections/figs/comp_gopro}

\input{sections/figs/tta_reds}

\input{sections/figs/comp_real}

%% file: sections/tables/tta_gopro_reds.tex
\begin{table}[!t]
    \vspace{-6mm}
    \centering
    \footnotesize
    \renewcommand{\arraystretch}{1.15}
    \newcommand{\justlone}{\multicolumn{1}{l |}{$\mathcal{L}_{1}$}}
    \newcommand{\plusreblur}[1]{\multicolumn{1}{l |}{$\mathcal{L}_{1} + \mathcal{L}_{\text{Reblur, n{#1}}}$}}
    
    \begin{tabularx}{\linewidth}{l p{2.2cm} >{\centering\arraybackslash}X >{\centering\arraybackslash}X >{\centering\arraybackslash}X >{\centering\arraybackslash}X >{\centering\arraybackslash}X >{\centering\arraybackslash}X >{\centering\arraybackslash}X >{\centering\arraybackslash}X}
        \toprule
        &  & \multicolumn{4}{c}{On GOPRO dataset} & \multicolumn{4}{c}{On REDS dataset} \\
        Model & \multicolumn{1}{c |}{Optimization} & LPIPS$_\downarrow$ & NIQE$_\downarrow$ & PSNR$^\uparrow$ & \multicolumn{1}{c |}{SSIM$^\uparrow$} &  LPIPS$_\downarrow$ & NIQE$_\downarrow$ & PSNR$^\uparrow$ & SSIM$^\uparrow$ \\
        \midrule
        \multirow{5}{*}{U-Net} & \justlone & 0.1635 & 5.996 & 29.66 & \multicolumn{1}{c |}{0.8874} & 0.1486 & 3.649 & 30.80 & 0.8772 \\
        & \plusreblur{1} & 0.1365 & 5.629 & 29.58 & \multicolumn{1}{c |}{0.8869} & 0.1435 & 3.487 & 30.76 & 0.8776 \\
        & \multicolumn{1}{r |}{$+$ TTA step 5} & \textbf{0.1327} & \textbf{5.599} & 29.52 & \multicolumn{1}{c |}{0.8878} & \textbf{0.1403} & \textbf{3.476} & 30.64 & 0.8781 \\
        & \plusreblur{2} & 0.1238 & 5.124 & 29.44 & \multicolumn{1}{c |}{0.8824} & 0.1252 & 2.918 & 30.46 & 0.8717 \\
        & \multicolumn{1}{r |}{$+$ TTA step 5} & \textbf{0.1187} & \textbf{5.000} & 29.42 & \multicolumn{1}{c |}{0.8831} & \textbf{0.1226} & \textbf{2.849} & 30.25 & 0.8701 \\
        \midrule
        \multirow{5}{*}{SRN} & \justlone & 0.1246 & 5.252 & 30.62 & \multicolumn{1}{c |}{0.9078} & 0.1148 & 3.392 & 31.89 & 0.8999 \\
        & \plusreblur{1} & 0.1140 & 5.136 & 30.74 & \multicolumn{1}{c |}{0.9104} & 0.1071 & 3.305 & 32.01 & 0.9044 \\
        & \multicolumn{1}{r |}{$+$ TTA step 5} & \textbf{0.1101} & \textbf{5.079} & 30.60 & \multicolumn{1}{c |}{0.9100} & \textbf{0.1029} & \textbf{3.278} & 31.83 & 0.9040 \\
        & \plusreblur{2} & 0.1037 & 4.887 & 30.57 & \multicolumn{1}{c |}{0.9074} & 0.0947 & 2.875 & 31.82 & 0.9026 \\
        & \multicolumn{1}{r |}{$+$ TTA step 5} & \textbf{0.0983} & \textbf{4.730} & 30.44 & \multicolumn{1}{c |}{0.9067} & \textbf{0.0909} & \textbf{2.798} & 31.50 & 0.9008 \\
        \midrule
        \multirow{5}{*}{DHN} & \justlone & 0.1179 & 5.490 & 31.53 & \multicolumn{1}{c |}{0.9207} & 0.0942 & 3.288 & 32.65 & 0.9152 \\
        & \plusreblur{1} & 0.0975 & 5.472 & 31.53 & \multicolumn{1}{c |}{0.9217} & 0.0931 & 3.248 & 32.57 & 0.9143 \\
        & \multicolumn{1}{r |}{$+$ TTA step 5} & \textbf{0.0940} & \textbf{5.343} & 31.32 & \multicolumn{1}{c |}{0.9208} & \textbf{0.0887} & \textbf{3.220} & 32.38 & 0.9139 \\
        & \plusreblur{2} & 0.0837 & 5.076 & 31.34 & \multicolumn{1}{c |}{0.9177} & 0.0805 & 2.830 & 32.44 & 0.9122 \\
        & \multicolumn{1}{r |}{$+$ TTA step 5} & \textbf{0.0805} & \textbf{4.948} & 31.28 & \multicolumn{1}{c |}{0.9174} & \textbf{0.0763} & \textbf{2.761} & 32.17 & 0.9110 \\
        \bottomrule
    \end{tabularx}
    \\
    \tabcspace
    \caption{
        \textbf{Quantitative analysis of the reblurring losses and test-time adaptation applied to various deblurring networks on GOPRO and REDS datasets.}
        }
    \label{tab:tta_comparison}
    \tabspace
\end{table}

%% file: sections/figs/loss_compare.tex
\begin{figure*}
    \renewcommand{\wp}{0.165\linewidth}
    \centering
    \subfloat{\includegraphics[width=\wp]{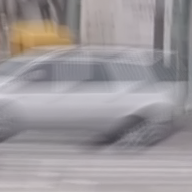}}
    \hfill
    \subfloat{\includegraphics[width=\wp]{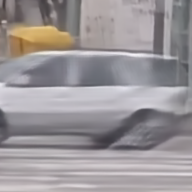}}
    \hfill
    \subfloat{\includegraphics[width=\wp]{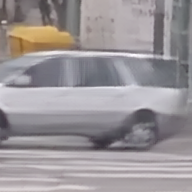}}
    \hfill
    \subfloat{\includegraphics[width=\wp]{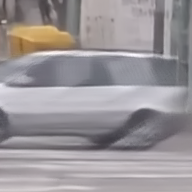}}
    \hfill
    \subfloat{\includegraphics[width=\wp]{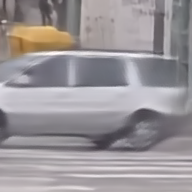}}
    \hfill
    \subfloat{\includegraphics[width=\wp]{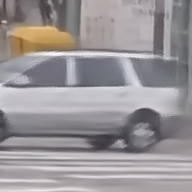}}
    \\
    \addtocounter{subfigure}{-6}
    \subfloat[$B$]{\includegraphics[width=\wp]{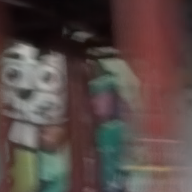}}
    \hfill
    \subfloat[$\mathcal{L}_1$]{\includegraphics[width=\wp]{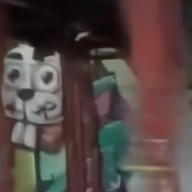}}
    \hfill
    \subfloat[$\mathcal{L}_\text{VGG}$]{\includegraphics[width=\wp]{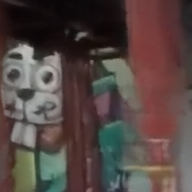}}
    \hfill
    \subfloat[$\mathcal{L}_\text{Adv}$]{\includegraphics[width=\wp]{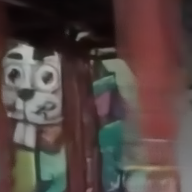}}
    \hfill
    \subfloat[$\mathcal{L}_\text{Reblur, n1}$]{\includegraphics[width=\wp]{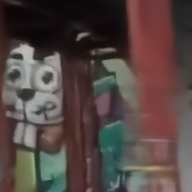}}
    \hfill
    \subfloat[$\mathcal{L}_\text{Reblur, n2}$]{\includegraphics[width=\wp]{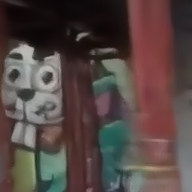}}
    \\
    \figcspace
    \caption{
        \textbf{Visual comparison of deblurred results by training loss function on GOPRO dataset.}
        \textbf{Upper:} SRN, \textbf{Lower:} U-Net.
    }
    \label{fig:fig_reblur_loss_gopro}
    \figspace
\end{figure*}

%% file: sections/tables/percep_loss_comp.tex
\begin{table}[t]
    \captionsetup{font=footnotesize}
    \footnotesize
    \renewcommand{\arraystretch}{1.15}
    \begin{minipage}[t]{0.49 \linewidth}
        \centering
        \begin{tabularx}{\linewidth}{p{2.2cm} >{\centering\arraybackslash}X  >{\centering\arraybackslash}X  >{\centering\arraybackslash}X  >{\centering\arraybackslash}X }
            \toprule
            Method & LPIPS$_\downarrow$ & NIQE$_\downarrow$ & PSNR$^\uparrow$ & SSIM$^\uparrow$ \\
            \midrule
            % U-Net (no $\mathcal{L}_{\text{R}}$) & 0.1635 & 5.996 & 29.66 & 0.8874\\
            \multicolumn{1}{l |}{U-Net (baseline)} & 0.1635 & 5.996 & 29.66 & 0.8874\\
            \multicolumn{1}{l |}{U-Net ($\mathcal{L}_\text{Blur}$)} & 0.1301 & 5.132 & 29.47 & 0.8839\\
            \multicolumn{1}{r |}{$+ \mathcal{L}_\text{Sharp}$ with $S$} & 0.1410 & 5.307 & 29.15 & 0.8694\\
            \multicolumn{1}{r |}{$+ \mathcal{L}_\text{Sharp}$ with $\hat{S}$} & \textbf{0.1238} & \textbf{5.124} & 29.44 & 0.8824\\
            \bottomrule
        \end{tabularx}
        \tabcspace
        \caption{
            \textbf{The effect of the sharpness preservation in training our reblurring module measured on GOPRO dataset.}
            %
            % In \eqref{eq:eq_loss_sp}, using the pseudo-sharp image $\hat{S}$ instead of the real one $S$ leads to better deblurring performance.
            % %
            % We note that the reblurring module is constructed using 2 ResBlocks.
        }
        \label{tab:gopro_sharpness_preservation}
    \end{minipage}
    \hfill
    \begin{minipage}[t]{0.49 \linewidth}
        \centering
        \begin{tabularx}{\linewidth}{p{2.2cm} >{\centering\arraybackslash}X  >{\centering\arraybackslash}X  >{\centering\arraybackslash}X  >{\centering\arraybackslash}X }
            \toprule
            Method & LPIPS$_\downarrow$ & NIQE$_\downarrow$ & PSNR$^\uparrow$ & SSIM$^\uparrow$ \\
            \midrule
            \multicolumn{1}{l |}{SRN ($\mathcal{L}_{1}$)} & 0.1246 & 5.252 & 30.62 & 0.9078\\
            \multicolumn{1}{r |}{$+ 0.001\mathcal{L}_{\text{Adv}}$} & 0.1141 & 4.960 & 30.53 & 0.9068\\
            \multicolumn{1}{r |}{$+ 0.3\mathcal{L}_{\text{VGG}}$} & \textbf{0.1037} & 4.945 & 30.60 & 0.9074\\
            \multicolumn{1}{r |}{$+ \mathcal{L}_{\text{Reblur, n2}}$} & \textbf{0.1037} & \textbf{4.887} & 30.57 & 0.9074\\
            \bottomrule
        \end{tabularx}
        \tabcspace
        \caption{
            % \textbf{Comparison of reblurring loss and other perceptual losses on GOPRO~\cite{Nah_2017_CVPR} dataset applied to SRN.}
            \textbf{Comparison of reblurring loss and other perceptual losses on GOPRO dataset applied to SRN.}
        }
        \label{tab:comp_reblur_adv}
    \end{minipage}
    \tabspace
\end{table}

%% file: sections/figs/comp_gopro.tex
\begin{figure}
    \centering
    \renewcommand{\wp}{0.248\linewidth}
    \newcommand{\wwp}{0.124\linewidth}
    \subfloat{\includegraphics[width=\wwp]{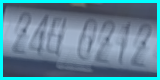}}
    \subfloat{\includegraphics[width=\wwp]{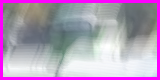}}
    \hfill
    \subfloat{\includegraphics[width=\wwp]{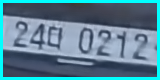}}
    \subfloat{\includegraphics[width=\wwp]{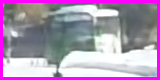}}
    \hfill
    \subfloat{\includegraphics[width=\wwp]{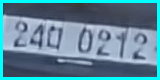}}
    \subfloat{\includegraphics[width=\wwp]{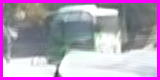}}
    \hfill
    \subfloat{\includegraphics[width=\wwp]{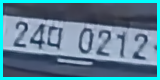}}
    \subfloat{\includegraphics[width=\wwp]{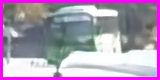}}
    \\
    \subfloat{\includegraphics[width=\wp]{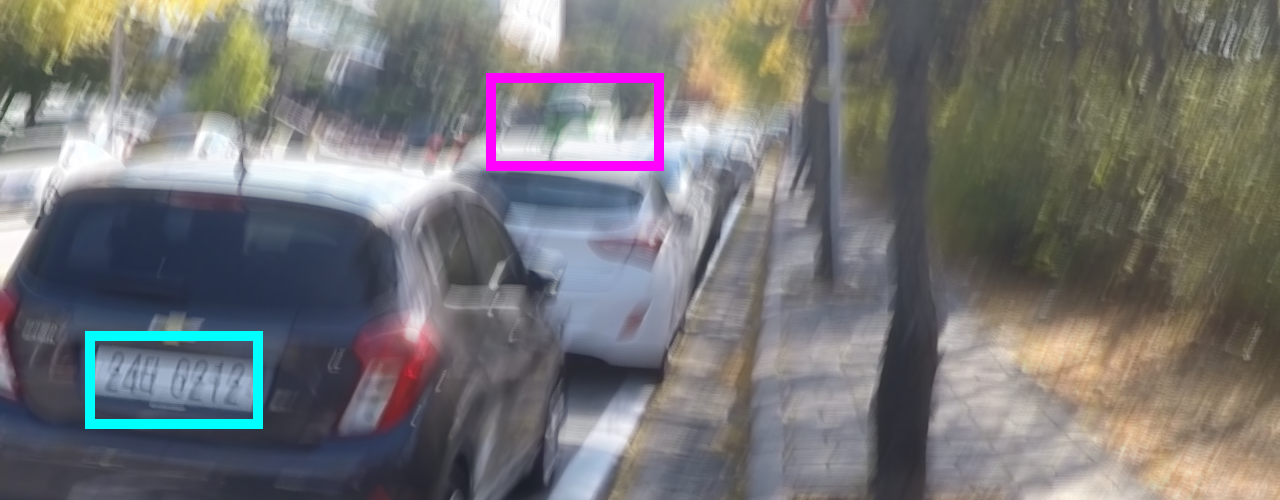}}
    \hfill
    \subfloat{\includegraphics[width=\wp]{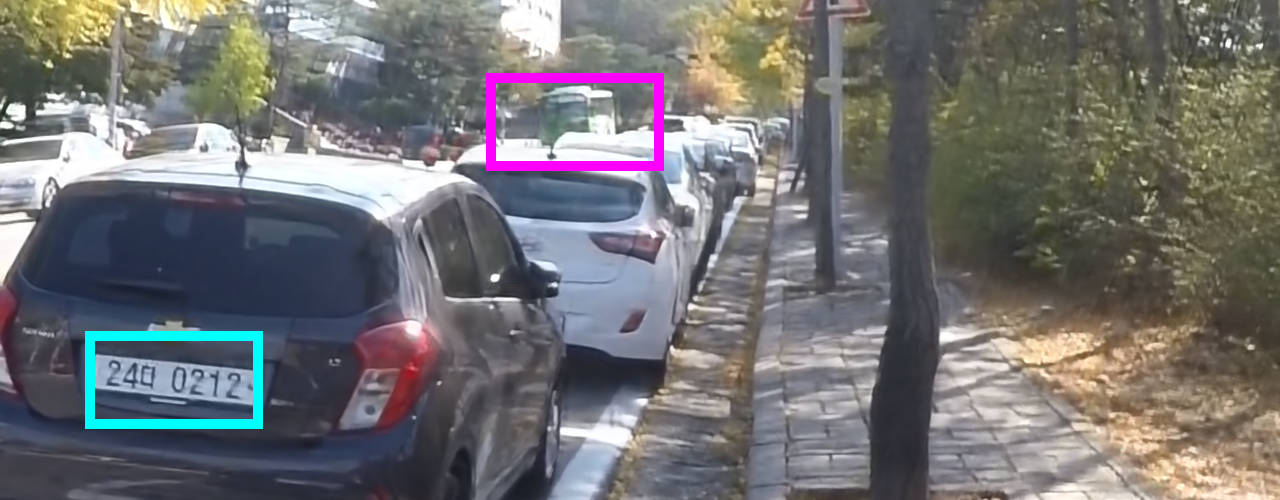}}
    \hfill
    \subfloat{\includegraphics[width=\wp]{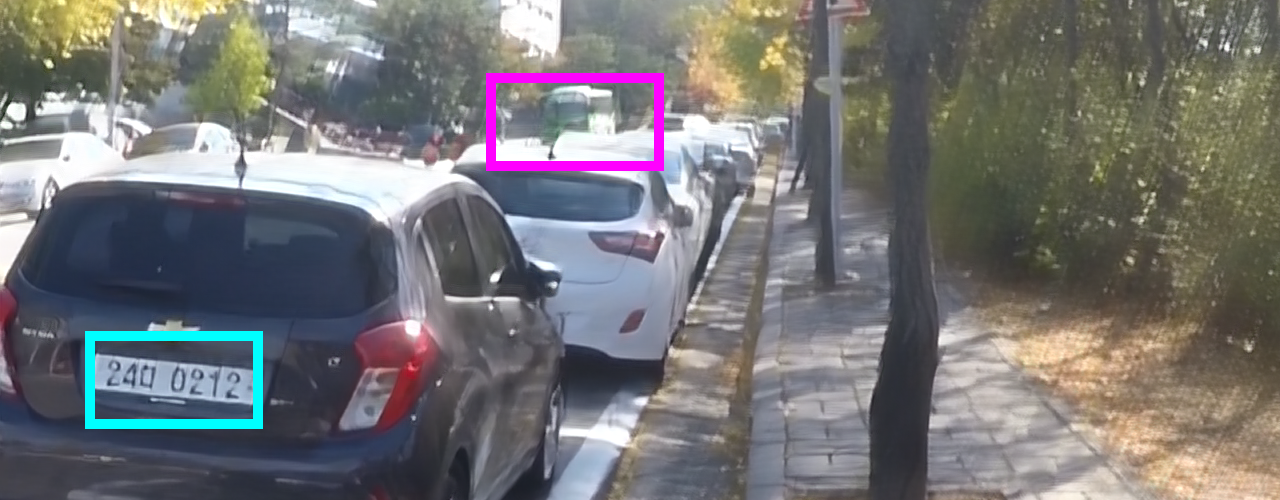}}
    \hfill
    \subfloat{\includegraphics[width=\wp]{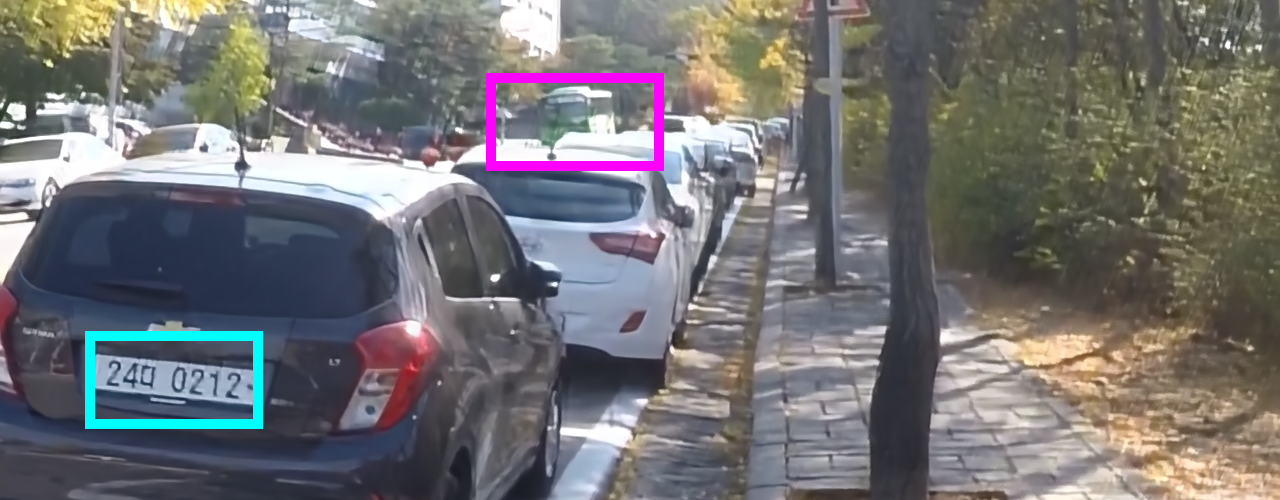}}
    \\
    \subfloat{\includegraphics[width=\wwp]{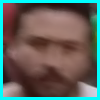}}
    \subfloat{\includegraphics[width=\wwp]{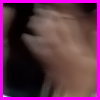}}
    \hfill
    \subfloat{\includegraphics[width=\wwp]{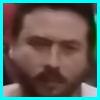}}
    \subfloat{\includegraphics[width=\wwp]{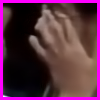}}
    \hfill
    \subfloat{\includegraphics[width=\wwp]{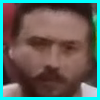}}
    \subfloat{\includegraphics[width=\wwp]{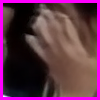}}
    \hfill
    \subfloat{\includegraphics[width=\wwp]{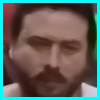}}
    \subfloat{\includegraphics[width=\wwp]{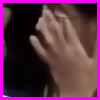}}
    \\
    \addtocounter{subfigure}{-20}
    \subfloat[Blurry input $B$]{\includegraphics[width=\wp]{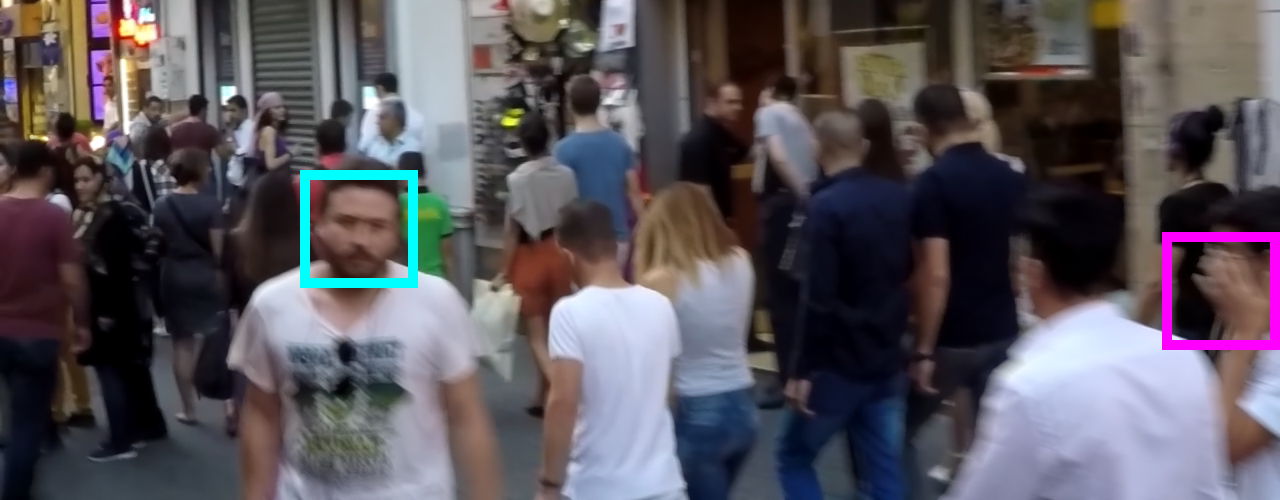}}
    \hfill
    \subfloat[SE-Sharing]{\includegraphics[width=\wp]{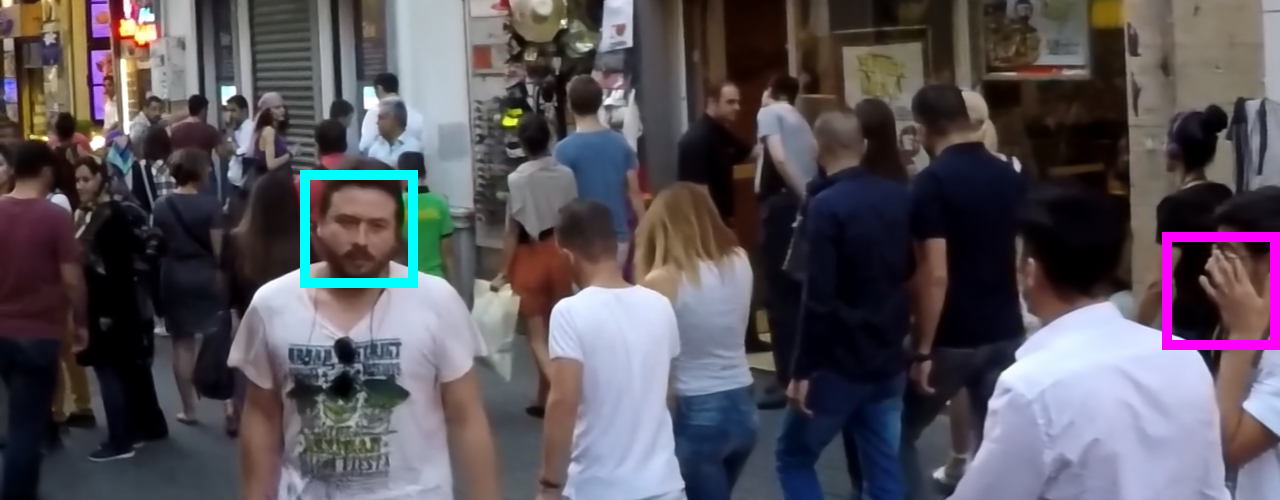}}
    \hfill
    \subfloat[DeblurGAN-v2]{\includegraphics[width=\wp]{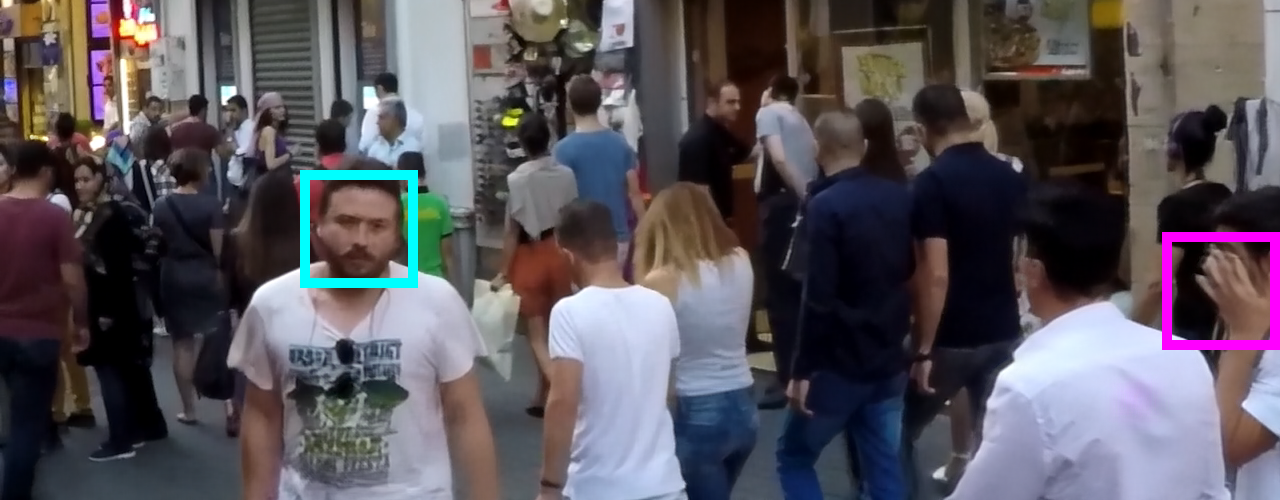}}
    \hfill
    \subfloat[\textbf{Ours (TTA step 5)}]{\includegraphics[width=\wp]{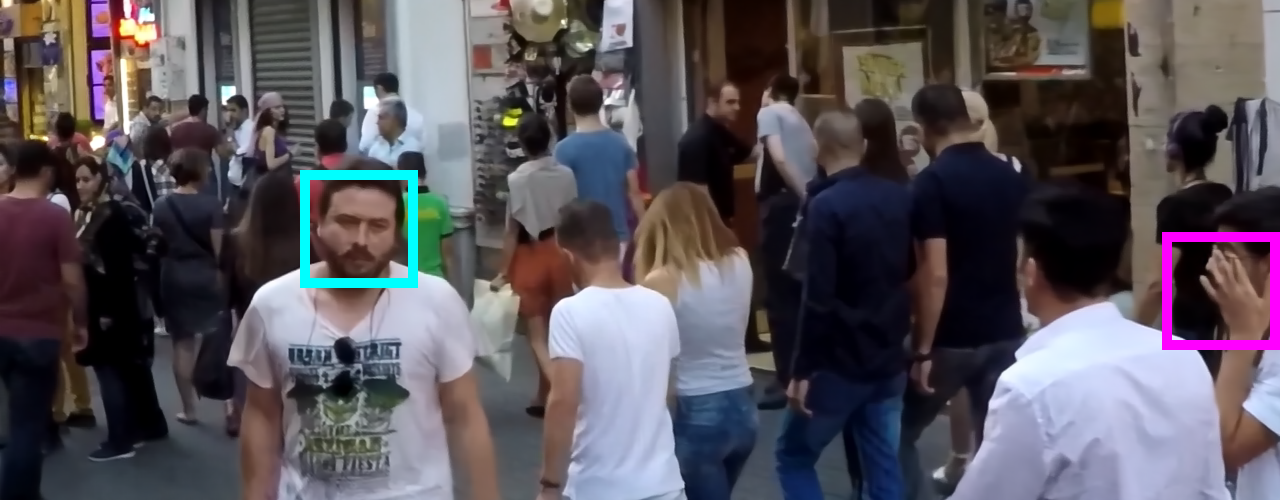}}
    \figcspace
    \caption{
        \textbf{Qualitative comparison between state-of-the-art deblurring methods on the GOPRO dataset.}
        We used the SRN model as a baseline architecture.
    }
    \label{fig:fig_gopro}
    \figspace
\end{figure}

%% file: sections/figs/tta_reds.tex
\begin{figure}[t]
    \vspace{-4mm}
    \captionsetup[subfloat]{font=small}
    \renewcommand{\wp}{0.245\linewidth}
    \centering
    \subfloat{\includegraphics[width=\wp]{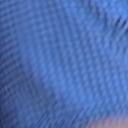}}
    \hfill
    \subfloat{\includegraphics[width=\wp]{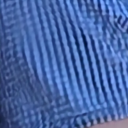}}
    \hfill
    \subfloat{\includegraphics[width=\wp]{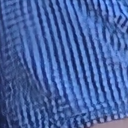}}
    \hfill
    % \subfloat{\includegraphics[width=\wp]{figs/tta_comparison/1_14_TTA_00000090.png}}
    % \subfloat{\includegraphics[width=\wp]{figs/tta_comparison/1_14_TTA10_00000090.png}}
    \subfloat{\includegraphics[width=\wp]{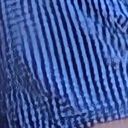}}
    \\
    % \subfloat{\includegraphics[width=\wp]{figs/tta_comparison/1_Blur_00000005.png}}
    % \hfill
    % \subfloat{\includegraphics[width=\wp]{figs/tta_comparison/1_L1_00000005.png}}
    % \hfill
    % \subfloat{\includegraphics[width=\wp]{figs/tta_comparison/1_Reblur_n2_00000005.png}}
    % \hfill
    % \subfloat{\includegraphics[width=\wp]{figs/tta_comparison/1_TTA_00000005.png}}
    % \\
    \addtocounter{subfigure}{-4}
    % \subfloat[$B$]{\includegraphics[width=\wp]{figs/tta_comparison/1_Blur_00000005.png}}
    % \hfill
    % \subfloat[$\mathcal{L}_1$]{\includegraphics[width=\wp]{figs/tta_comparison/1_L1_00000005.png}}
    % \hfill
    % \subfloat[$\mathcal{L}_\text{Reblur, n2}$]{\includegraphics[width=\wp]{figs/tta_comparison/1_Reblur_n2_00000005.png}}
    % \hfill
    % \subfloat[$+$ TTA step 5]{\includegraphics[width=\wp]{figs/tta_comparison/1_TTA_00000005.png}}
    % \hfill
    \subfloat[$B$]{\includegraphics[width=\wp]{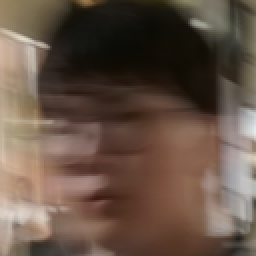}}
    \hfill
    \subfloat[$\mathcal{L}_1$]{\includegraphics[width=\wp]{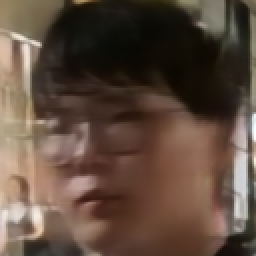}}
    \hfill
    \subfloat[$\mathcal{L}_\text{Reblur, n2}$]{\includegraphics[width=\wp]{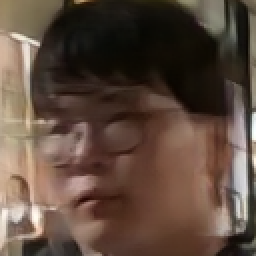}}
    \hfill
    % \subfloat[$+$ TTA step 5]{\includegraphics[width=\wp]{figs/tta_comparison/1_TTA_00000063.png}}
    % \subfloat[$+$ TTA step 10]{\includegraphics[width=\wp]{figs/tta_comparison/1_TTA10_00000063.png}}
    \subfloat[$+$ TTA step 20]{\includegraphics[width=\wp]{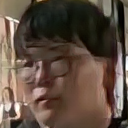}}
    \figcspace
    \caption{
        \textbf{Qualitative comparison between different training objectives and the test-time adaptation.}
        Patches are sampled from the REDS dataset validation set.
    }
    \label{fig:fig_tta_reds}
    \figspace
    \vspace{2mm}
\end{figure}

%% file: sections/figs/comp_real.tex
\begin{figure*}
    \centering
    \renewcommand{\wp}{0.248\linewidth}
    \newcommand{\wwp}{0.124\linewidth}
    \subfloat{\includegraphics[width=\wwp]{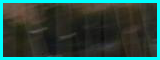}}
    \subfloat{\includegraphics[width=\wwp]{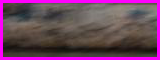}}
    \hfill
    \subfloat{\includegraphics[width=\wwp]{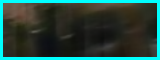}}
    \subfloat{\includegraphics[width=\wwp]{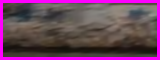}}
    \hfill
    \subfloat{\includegraphics[width=\wwp]{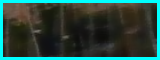}}
    \subfloat{\includegraphics[width=\wwp]{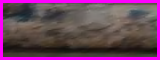}}
    \hfill
    \subfloat{\includegraphics[width=\wwp]{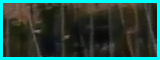}}
    \subfloat{\includegraphics[width=\wwp]{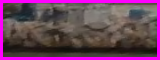}}
    \\
    \subfloat{\includegraphics[width=\wp]{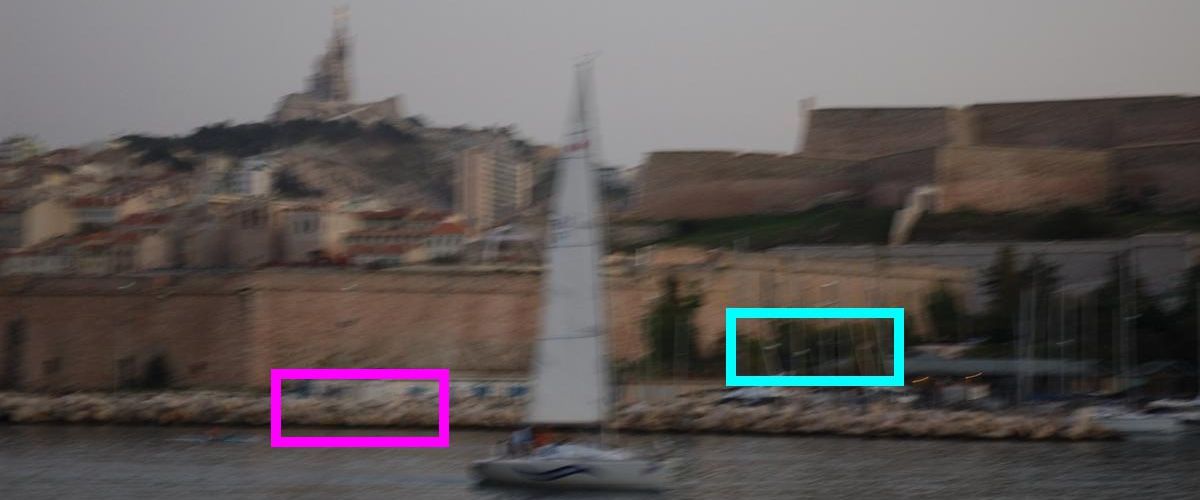}}
    \hfill
    \subfloat{\includegraphics[width=\wp]{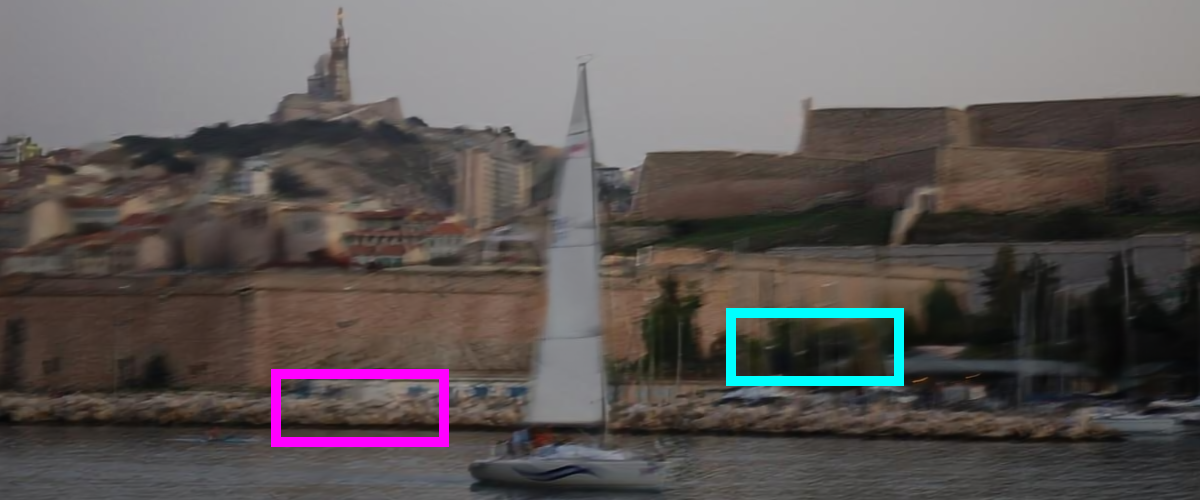}}
    \hfill
    \subfloat{\includegraphics[width=\wp]{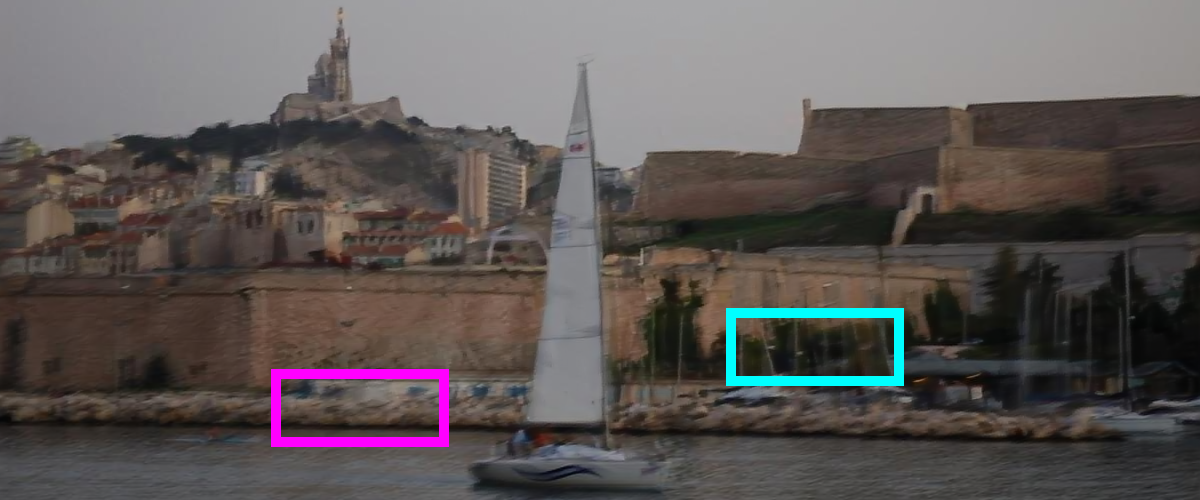}}
    \hfill
    \subfloat{\includegraphics[width=\wp]{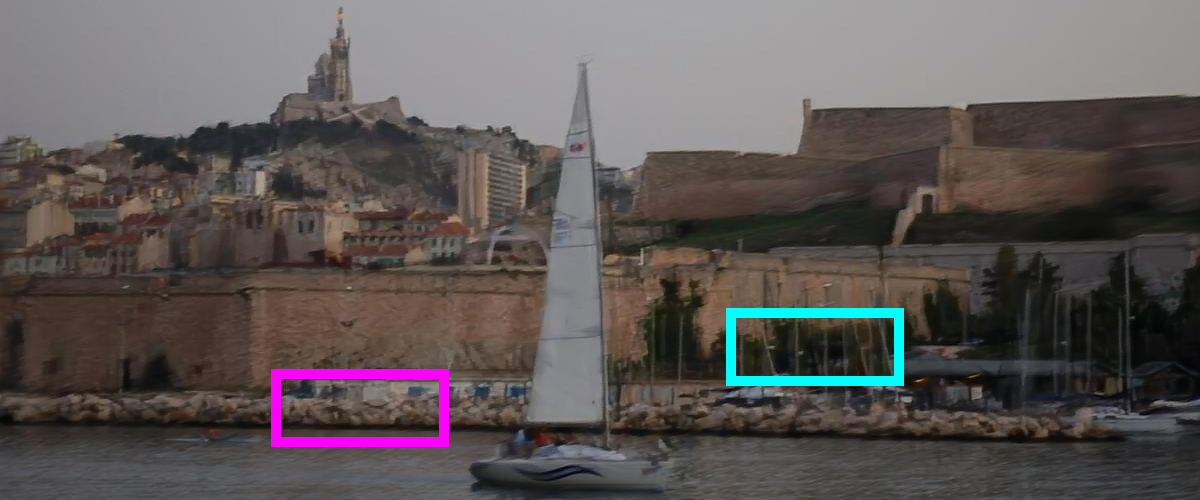}}
    \\
    \subfloat{\includegraphics[width=\wwp]{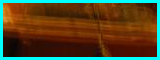}}
    \subfloat{\includegraphics[width=\wwp]{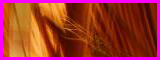}}
    \hfill
    \subfloat{\includegraphics[width=\wwp]{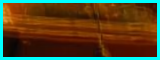}}
    \subfloat{\includegraphics[width=\wwp]{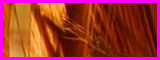}}
    \hfill
    \subfloat{\includegraphics[width=\wwp]{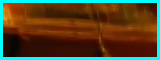}}
    \subfloat{\includegraphics[width=\wwp]{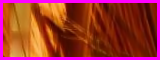}}
    \hfill
    \subfloat{\includegraphics[width=\wwp]{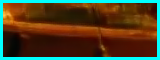}}
    \subfloat{\includegraphics[width=\wwp]{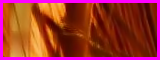}}
    \\
    \subfloat{\includegraphics[width=\wp]{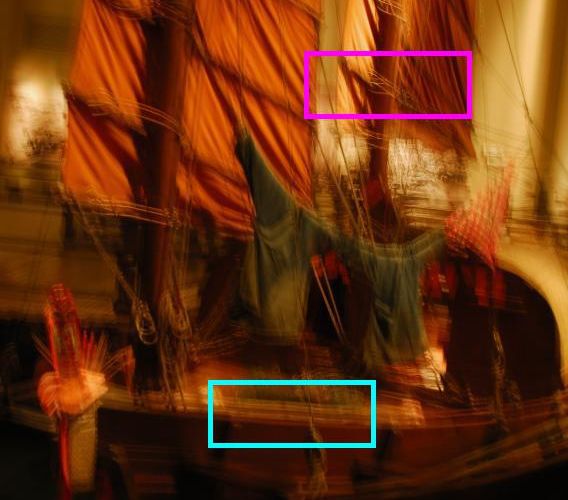}}
    \hfill
    \subfloat{\includegraphics[width=\wp]{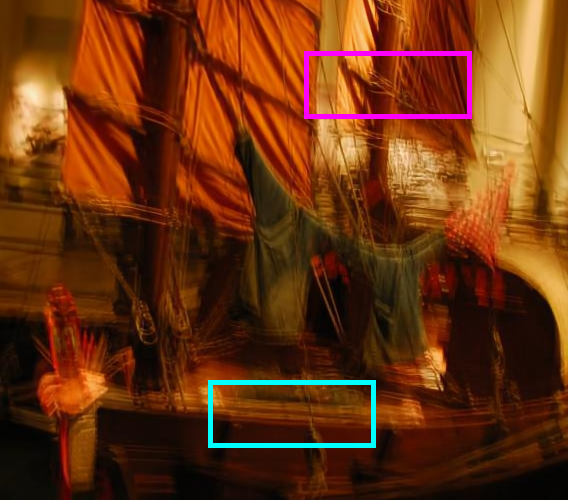}}
    \hfill
    \subfloat{\includegraphics[width=\wp]{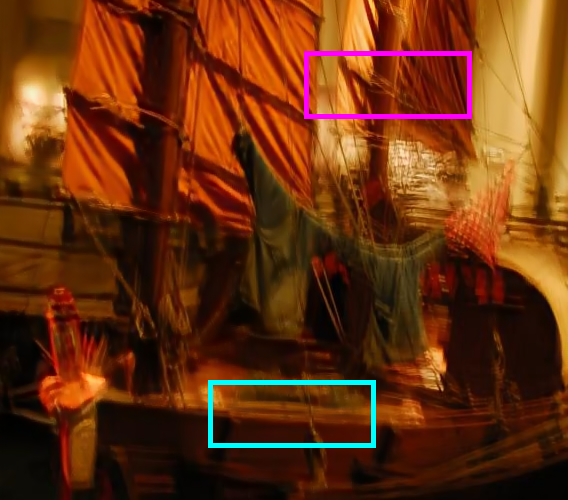}}
    \hfill
    \subfloat{\includegraphics[width=\wp]{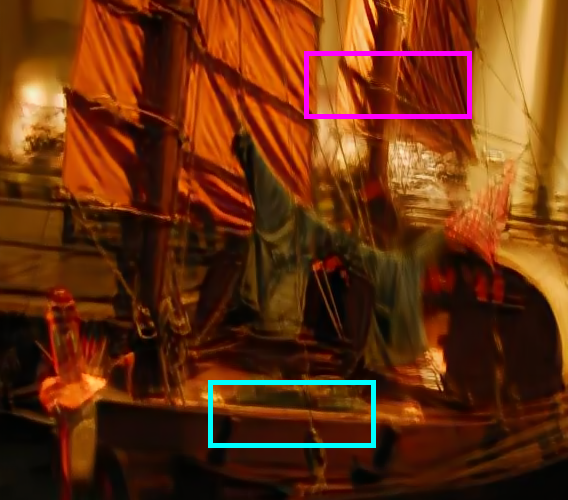}}
    \\
    \subfloat{\includegraphics[width=\wwp]{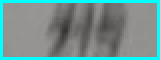}}
    \subfloat{\includegraphics[width=\wwp]{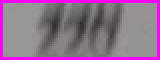}}
    \hfill
    \subfloat{\includegraphics[width=\wwp]{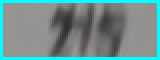}}
    \subfloat{\includegraphics[width=\wwp]{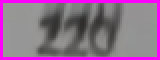}}
    \hfill
    \subfloat{\includegraphics[width=\wwp]{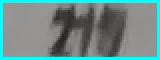}}
    \subfloat{\includegraphics[width=\wwp]{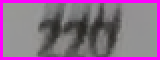}}
    \hfill
    \subfloat{\includegraphics[width=\wwp]{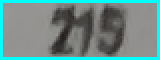}}
    \subfloat{\includegraphics[width=\wwp]{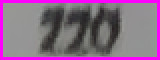}}
    \\
    \addtocounter{subfigure}{-32}
    \subfloat[Blurry input $B$]{\includegraphics[width=\wp]{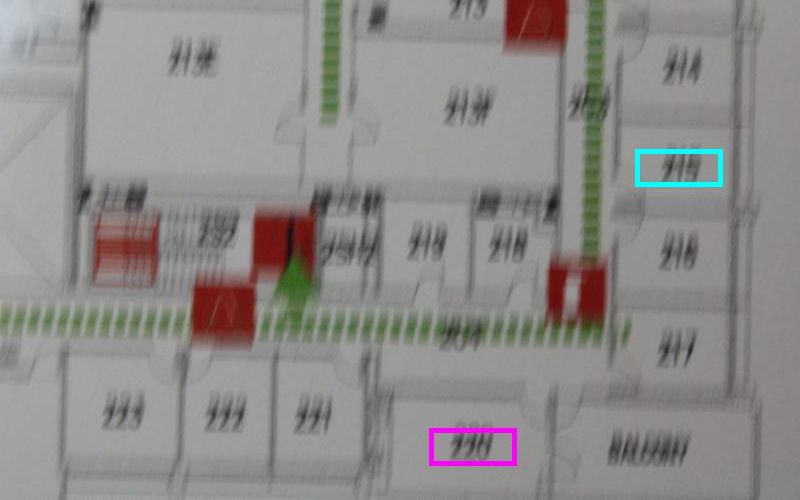}}
    \hfill
    \subfloat[$\mathcal{L}_1$]{\includegraphics[width=\wp]{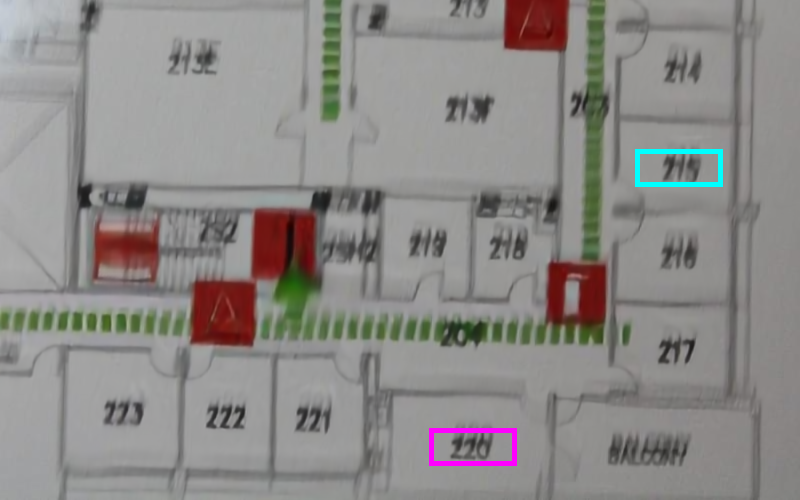}}
    \hfill
    \subfloat[\textbf{$\mathcal{L}_\text{Reblur, n2}$ (Ours)}]{\includegraphics[width=\wp]{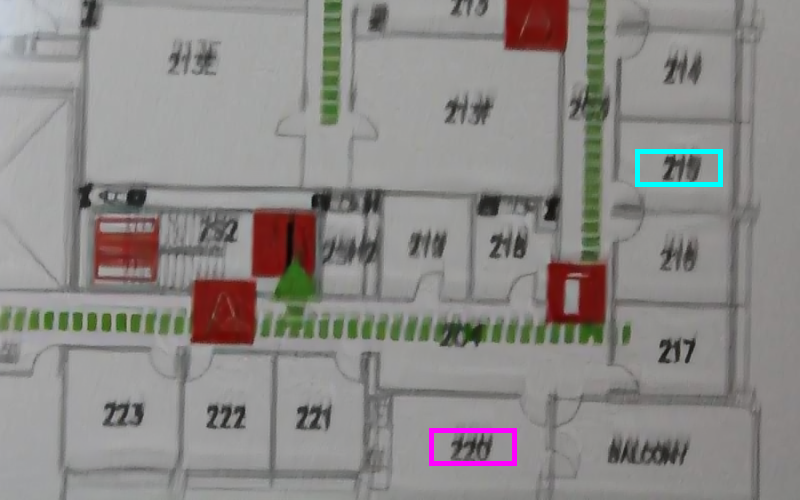}}
    \hfill
    \subfloat[\textbf{$+$ TTA step 5 (Ours)}]{\includegraphics[width=\wp]{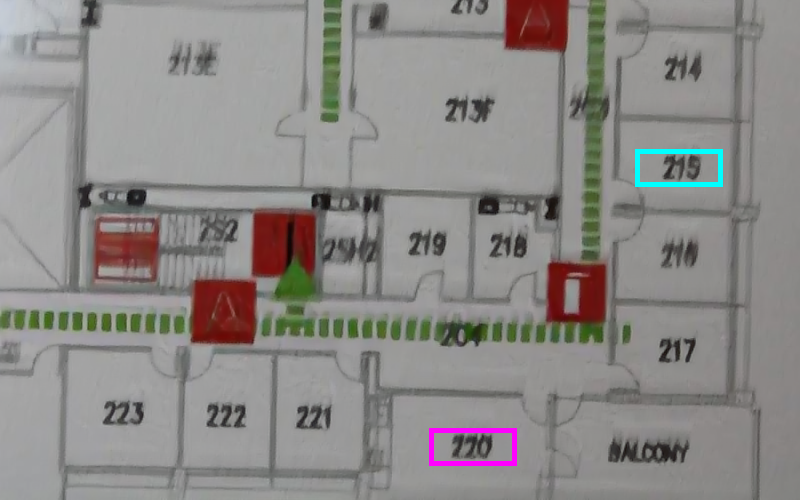}}
    \figcspace
    \caption{
        \textbf{Qualitative comparison of deblurring results on the real-world images~\citep{Lai_2016_CVPR} by different loss functions and test-time adaptation.}
        The proposed test-time adaptation greatly improves visual quality and sharpness of the deblurred images.
    }
    \label{fig:fig_real}
    \figspace
\end{figure*}

%% file: sections/camera_ready/5_conclusion.tex
\section{Conclusion}
\label{sec:conclusion}

In this paper, we validate a new observation that clean sharp images are hard to reblur and develop novel low-level perceptual objective terms, namely reblurring loss.
The term is constructed to care the image blurriness by jointly training a pair of deblurring and reblurring modules.
The supervised reblurring loss provides an amplified comparison on motion blur while the self-supervised loss inspects the blurriness in a single image with the learned reblurring module.
The self-supervision lets the deblurring module adapt to the new image at test time without ground truth.
By applying the loss terms to state-of-the-art deblurring methods, we demonstrate our method consistently improves the the perceptual sharpness of the deblurred images visually as well as quantitatively.

%% file: sections/camera_ready/aa_acknowledgment.tex
\subsubsection*{Acknowledgments}

% This work was supported in part by IITP grant funded by the Korea government [No. 2021-0-01343, Artificial Intelligence Graduate School Program (Seoul National University)], and in part by Hyundai Motor Group through HMG-SNU AI Consortium fund.

This work was supported in part by IITP grant funded by the Korea government [No. 2021-0-01343, Artificial Intelligence Graduate School Program (Seoul National University)], and in part by AIRS Company in Hyundai Motor Company \& Kia Motors Corporation through HMC/KIA-SNU AI Consortium.

%% file: sections/camera_ready/a_appendix.tex
\clearpage

\setcounter{figure}{0}
\setcounter{table}{0}

\renewcommand{\thetable}{\Alph{table}}
\renewcommand{\thefigure}{\Alph{figure}}
\renewcommand{\theequation}{\Alph{equation}}
\renewcommand{\thealgorithm}{\Alph{algorithm}}

\section{Appendix}
\label{sec:appendix}
In this appendix, we provide the implementation details and additional experimental analysis.
% explain the detailed experimental results that are not shown in the main manuscript.
In Section~\ref{sec:implementation}, we explain the implementation details with the model architecture specifics, training details, and the evaluation metrics.
Section~\ref{sec:reblur_module_size} describes how the reblurring module design and the size are determined.
Then in Section~\ref{sec:combination}, we describe the different characteristics of the proposed reblurring loss and the other perceptual losses.
We combine our reblurring loss with the other perceptual losses to take advantage in multiple perspectives.
In Section \ref{sec:implementation}, the test-time adaptation algorithm is described.
In Section~\ref{sec:trade_off}, we show the quantitative effect of test-time adaptation and show the trade-off relation between the conventional distortion quality metric (PSNR, SSIM) and the perceptual metrics (LPIPS, NIQE) compared with the baselines.
% In section~\ref{sec:more_examples}, we visually validate the effect of reblurring loss and test-time adaptation.

% \section{Reblurring Concept}

\section{Implementation Details}
\label{sec:implementation}

\noindent \textbf{Model Architecture.}
In the main manuscript, we mainly performed the experiments with 3 different model architectures.
First, we set our baseline model as a light-weight residual U-Net architecture that runs in a fast speed.
The baseline model is used to design our reblurring loss with pseudo-sharp images through ablation study in Table~\ref{tab:gopro_sharpness_preservation}.

\begin{figure}[h]
    \centering
    \includegraphics[width=\linewidth]{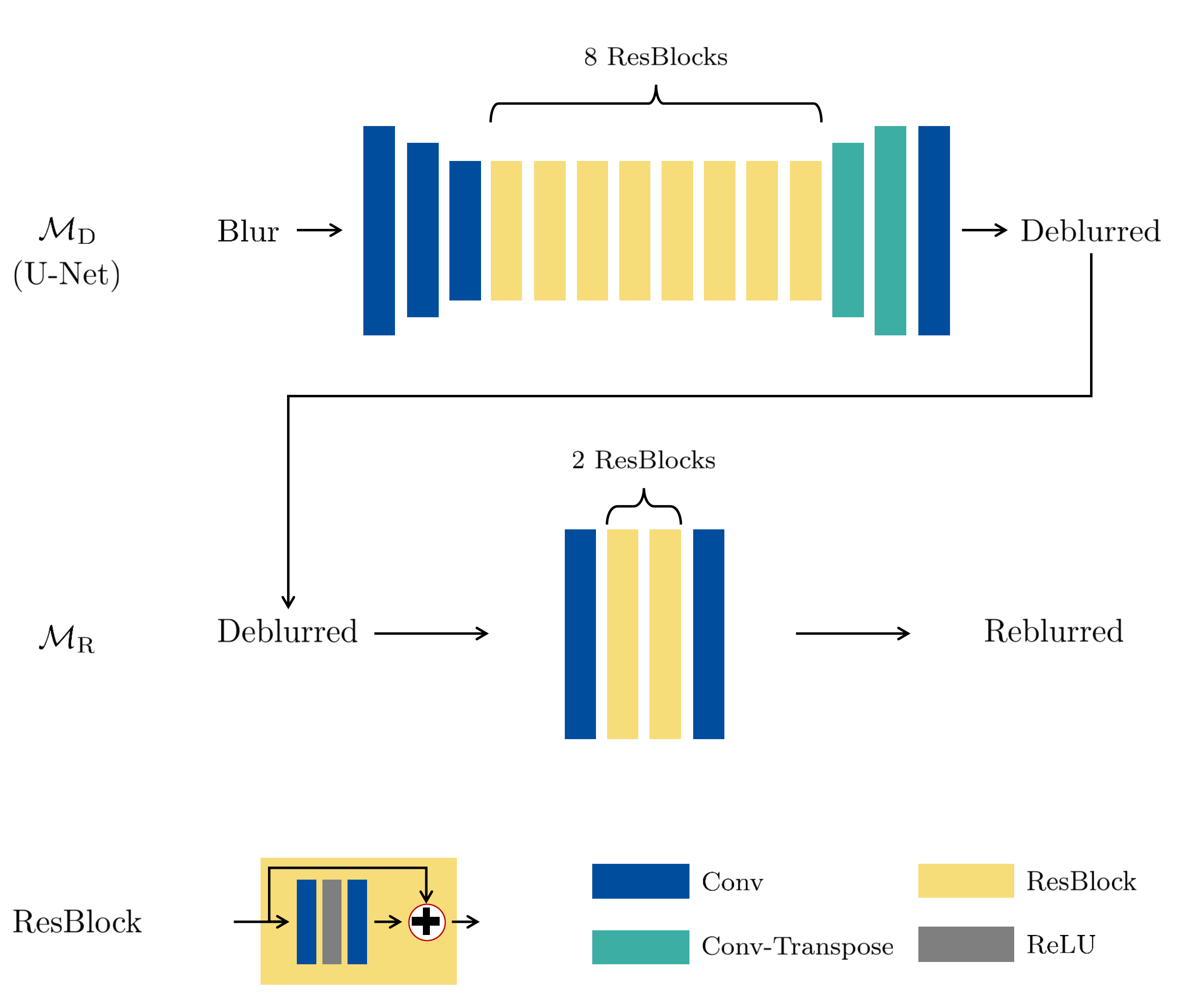}
    \\
    \figcspace
    \caption{\textbf{The baseline U-Net architecture and the reblurring module architecture}
    We use the same reblurring module for all experiments except the number of ResBlocks.
    }
    \label{fig:model_details}
    \figspace
\end{figure}

For reblurring operation, we use a simple residual network $\mathcal{M}_{\text{R}}$ without strides to avoid deconvolution artifacts.
The baseline U-Net and the reblurring module architectures are shown in Figure~\ref{fig:model_details}.
The detailed parameters for U-Net and $\mathcal{M}_{\text{R}}$ are each specified in Table~\ref{tab:unet_specs} and \ref{tab:reblur_specs}.

\begin{table}[t]
    \centering
    \small
    \renewcommand{\arraystretch}{1.15}
    \newcolumntype{P}[1]{>{\centering\arraybackslash}p{#1}}
    \newcolumntype{Y}{>{\centering\arraybackslash}X}
    
    % \begin{tabularx}{\linewidth}{Y Y Y}
    \begin{tabularx}{0.7 \linewidth}{P{1cm} Y Y}
        \toprule
        \# & Layer description & Output shape \\
        \midrule
        & Input & $3\times H \times W$ \\
        
        1 & $5\times5$ conv & $64 \times H \times W$\\
        2 & $3\times3$ conv & $128 \times H/2 \times W/2$\\
        3 & $3\times3$ conv & $192 \times H/4 \times W/4$\\
        
        4-19 & 8 ResBlocks ($3\times3$) & $192 \times H/4 \times W/4$\\
        
        20 & $3\times3$ conv & $128\times H/2\times W/2$\\
        21 & $3\times3$ conv & $64 \times H \times W$\\
        22 & $5\times5$ conv & $3 \times H \times W$\\
        \bottomrule
    \end{tabularx}
    \tabcspace
    \caption{\textbf{U-Net module specifics}}
    \label{tab:unet_specs}
    \tabspace
\end{table}

In addition to the U-Net, experiments were conducted with state-of-the-art deblurring models based on SRN~\citep{Tao_2017_ICCV} and DMPHN~\citep{Zhang_2019_CVPR_DMPHN}.
SRN~\citep{Tao_2018_CVPR} was originally designed to operate on grayscale images with a LSTM module.
Later, the authors released the sRGB version code without LSTM, exhibiting an improved accuracy.
We adopted the revised SRN structure in our experiments.

The other model we chose is based on DMPHN~(1-2-4-8)~\citep{Zhang_2019_CVPR_DMPHN}.
DMPHN performs hierarchical residual refinement to produce the final output.
The model consists of convolutional layers with ReLU activations that are spatially shift-equivariant.
In \citet{Zhang_2019_CVPR_DMPHN}, each level splits the given image and performs the convolutional operation on the divided patches.
As the convolutional weights do not differ by the input patches, the operations do not necessarily have to be done patch-wise.
Thus, we remove the multi-patch strategy and perform the convolution on the whole given input without dividing the image into patches.
We refer to the modified model as DHN.
As shown in Table~\ref{tab:dhn}, convolution on the whole image compared with patch-wise convolution brings higher accuracy.

\begin{table}[t]
    \centering
    \small
    \renewcommand{\arraystretch}{1.15}
    \newcolumntype{P}[1]{>{\centering\arraybackslash}p{#1}}
    \newcolumntype{Y}{>{\centering\arraybackslash}X}
    
    \begin{tabularx}{\linewidth}{p{3cm} Y Y Y Y }
        \toprule
        Method & LPIPS$_\downarrow$ & NIQE$_\downarrow$ & PSNR$^\uparrow$ & SSIM$^\uparrow$ \\
        \midrule
        DMPHN ($\mathcal{L}_{1}$ only) & 0.1184 & 5.542 & 31.42 & 0.9191\\
        DHN ($\mathcal{L}_{1}$ only) & \textbf{0.1179} & \textbf{5.490} & \textbf{31.53} & \textbf{0.9207}\\
        \bottomrule
    \end{tabularx}
    \tabcspace
    \caption{
        \textbf{DMPHN modification results on GOPRO dataset.}
        DHN without patch-wise convolution brings improved accuracy.
    }
    \label{tab:dhn}
    \tabspace
\end{table}

\noindent \textbf{Metrics.}
To quantitatively compare the deblurred images in the following sections, we use PSNR, SSIM~\citep{wang2004image}, LPIPS~\citep{Zhang_2018_CVPR_unreasonable}, and NIQE~\citep{mittal2012making}.
In the image deblurring literature, SSIM has been measured by MATLAB {\fontfamily{pcr}\selectfont ssim} function on sRGB images with $H\times W\times C$.
SSIM was originally developed for grayscale images and MATLAB {\fontfamily{pcr}\selectfont ssim} function for a 3-dimensional tensor considers an image to be a 3D grayscale volume image.
Thus, most of the previous SSIM measures were not accurate, leading to higher values.
Instead, we measured all the SSIM for each channel separately and averaged them.
We used {\fontfamily{pcr}\selectfont skimage.metrics.structural\_similarity} function in the scikit-image package for python to measure SSIM for multi-channel images.

\noindent \textbf{Training.}
For all the experiments, we performed the same training process for a fair comparison.
On the GOPRO dataset~\citep{Nah_2017_CVPR}, we trained each model for 4000 epochs.
On the REDS dataset~\citep{Nah_2019_CVPR_Workshops_REDS}, the models are trained for 200 epochs.
Adam~\citep{kingma2014adam} optimizer is used in all cases.
When calculating distance between images with Lp norm, we always set $p=1$, using L1 distance.
Starting from the initial learning rate $1\times10^{-4}$, the learning rate halves when training reaches 50\%, 75\%, and 90\% of the total epochs.
We used PyTorch 1.8.1 with CUDA 11.1 to implement the deblurring methods.
Mixed-precision training~\citep{micikevicius2017mixed} is employed to accelerate operations on RTX 2080 Ti GPUs.

\begin{table}[t]
    \centering
    \small
    \renewcommand{\arraystretch}{1.15}
    \newcolumntype{P}[1]{>{\centering\arraybackslash}p{#1}}
    \newcolumntype{Y}{>{\centering\arraybackslash}X}
    
    \begin{tabularx}{0.7 \linewidth}{P{1cm} Y Y}
        \toprule
        \# & Layer description & Output shape \\
        \midrule
        & Input & $3\times H \times W$ \\
        
        1 & $5\times5$ conv & $64 \times H \times W$\\
        2-5 & 2 ResBlocks ($5\times5$) & $64 \times H \times W$\\
        6 & $5\times5$ conv & $3 \times H \times W$\\
        \bottomrule
    \end{tabularx}
    \tabcspace
    \caption{\textbf{Reblurring module specifics}}
    \label{tab:reblur_specs}
    % \tabspace
\end{table}

\section{Determining the Reblurring Module Size}
\label{sec:reblur_module_size}

As our reblurring loss $\mathcal{L}_{\text{R}}$ is realized by $\mathcal{M}_{\text{R}}$, the reblurring module design plays an essential role.
As shown in Figure~\ref{fig:model_details}, the $\mathcal{M}_{\text{R}}$ architecture is a simple ResNet.
Table~\ref{tab:gopro_reblur_n} shows the relation between the deblurring performance and $\mathcal{M}_{\text{R}}$ size by changing the number of ResBlocks.

For all deblurring module $\mathcal{M}_{\text{D}}$ architectures, LPIPS was the best when the number of ResBlocks, $n=2$.
NIQE showed good performance when $2 \leq n \leq 3$.
PSNR and SSIM had tendency to decrease when $n \ge 1$.
For larger number of ResBlocks, we witnessed sharper edges could be obtained but sometimes, cartoon artifacts with over-strong edges were witnessed.

Considering the trade-off between the PSNR and the perceptual metrics, we chose $n \in \{1, 2\}$ in the following experiments.
$n=1$ finds balance between the PSNR and LPIPS and $n=2$ puts more weight on the perceptual quality.

\begin{table}[t]
    \centering
    \small
    \renewcommand{\arraystretch}{1.15}
    \newcolumntype{Y}{>{\centering\arraybackslash}X}
    
    \newcommand{\plusreblur}[1]{\multicolumn{1}{r|}{$+ \mathcal{L}_{\text{Reblur, n{#1}}}$}}
    
    \begin{tabularx}{\linewidth}{p{3cm} | Y Y Y Y}
        \toprule
        Method & LPIPS$_\downarrow$ & NIQE$_\downarrow$ & PSNR$^\uparrow$ & SSIM$^\uparrow$ \\
        \midrule
        U-Net ($\mathcal{L}_{1}$ only) & 0.1635 & 5.996 & \textbf{29.66} & \textbf{0.8874}\\
        \plusreblur{1} & 0.1365 & 5.629 & 29.58 & 0.8869\\
        \plusreblur{2} & \textbf{0.1238} & \textbf{5.124} & 29.44 & 0.8824\\
        \plusreblur{3} & 0.1386 & 5.448 & 29.38 & 0.8819\\
        \plusreblur{4} & 0.1415 & 5.513 & 29.25 & 0.8789\\
        \midrule
        SRN ($\mathcal{L}_{1}$ only) & 0.1246 & 5.252 & 30.62 & 0.9078\\
        \plusreblur{1} & 0.1140 & 5.136 & \textbf{30.74} & \textbf{0.9104}\\
        \plusreblur{2} & \textbf{0.1037} & 4.887 & 30.57 & 0.9074\\
        \plusreblur{3} & 0.1091 & \textbf{4.875} & 30.50 & 0.9060\\
        \plusreblur{4} & 0.1155 & 5.041 & 30.53 & 0.9056\\
        \midrule
        DHN ($\mathcal{L}_{1}$ only) & 0.1179 & 5.490 & \textbf{31.53} & 0.9207\\
        \plusreblur{1} & 0.0975 & 5.472 & \textbf{31.53} & \textbf{0.9217}\\
        \plusreblur{2} & \textbf{0.0837} & 5.076 & 31.34 & 0.9177\\
        \plusreblur{3} & 0.0845 & \textbf{4.963} & 31.26 & 0.9159\\
        \plusreblur{4} & 0.0861 & 5.041 & 31.19 & 0.9149\\
        \bottomrule
    \end{tabularx}
    \tabcspace
    \caption{
        \textbf{The effect of reblurring loss on GOPRO dataset by the reblurrimg module size.}
        Reblurring module size varies by the number of ResBlocks.
    }
    \label{tab:gopro_reblur_n}
    % \tabspace
\end{table}

\section{Combining Reblurring Loss with Other Perceptual Losses}
\label{sec:combination}

\begin{table}[t]
    \centering
    \small
    \begin{tabularx}{\linewidth}{p{3cm} | >{\centering\arraybackslash}X  >{\centering\arraybackslash}X  >{\centering\arraybackslash}X  >{\centering\arraybackslash}X }
        \toprule
        Method & LPIPS$_\downarrow$ & NIQE$_\downarrow$ & PSNR$^\uparrow$ & SSIM$^\uparrow$ \\
        \midrule
        SRN ($\mathcal{L}_{1}$ only) & 0.1246 & 5.252 & 30.62 & 0.9078\\
        % \multicolumn{1}{r|}{$+ \mathcal{L}_{\text{Reblur, n2}}$} & 0.1037 & 4.887 & 30.57 & 0.9074\\
        \multicolumn{1}{r|}{$+ \mathcal{L}_{\text{VGG}}$} & 0.1037 & 4.945 & 30.60 & 0.9074 \\
        \multicolumn{1}{r|}{$+ \mathcal{L}_{\text{VGG}} + \mathcal{L}_{\text{Reblur, n2}}$} & \textbf{0.0928} & \textbf{4.671} & 30.64 & 0.9079\\
        \multicolumn{1}{r|}{$+ \mathcal{L}_{\text{Adv}}$} & 0.1141 & 4.960 & 30.53 & 0.9068\\
        \multicolumn{1}{r|}{$+ \mathcal{L}_{\text{Adv}} + \mathcal{L}_{\text{Reblur, n2}}$} & \textbf{0.1014} & \textbf{4.811} & 30.56 & 0.9075\\
        \midrule
        DHN ($\mathcal{L}_{1}$ only) & 0.1179 & 5.490 & 31.53 & 0.9207\\
        \multicolumn{1}{r|}{$+ \mathcal{L}_{\text{VGG}}$} & 0.0994 & 5.022 & 31.48 & 0.9195\\
        \multicolumn{1}{r|}{$+ \mathcal{L}_{\text{VGG}} + \mathcal{L}_{\text{Reblur, n2}}$} & \textbf{0.0773} & \textbf{4.897} & 31.28 & 0.9161\\
        \multicolumn{1}{r|}{$+ \mathcal{L}_{\text{Adv}}$} & 0.0969 & 5.026 & 31.46 & 0.9188\\
        \multicolumn{1}{r|}{$+ \mathcal{L}_{\text{Adv}} + \mathcal{L}_{\text{Reblur, n2}}$} & \textbf{0.0835} & \textbf{4.799} & 31.28 & 0.9162\\
        \bottomrule
    \end{tabularx}
    \\
    \tabcspace
    \caption{
        % \textbf{Results on GOPRO~\cite{Nah_2017_CVPR} dataset from combined perceptual losses.}
        \textbf{Results on GOPRO dataset by adding reblurring loss to the other preceptual losses.}
        % Coefficients are omitted for better readability.
    }
    \label{tab:gopro_vgg_reblur}
    % \tabspace
\end{table}

\begin{table}[t]
    \centering
    \small
    \begin{tabularx}{\linewidth}{p{3cm} | >{\centering\arraybackslash}X  >{\centering\arraybackslash}X  >{\centering\arraybackslash}X  >{\centering\arraybackslash}X }
        \toprule
        Method & LPIPS$_\downarrow$ & NIQE$_\downarrow$ & PSNR$^\uparrow$ & SSIM$^\uparrow$ \\
        \midrule
        SRN ($\mathcal{L}_{1}$ only) & 0.1148 & 3.392 & 31.89 & 0.8999\\
        \multicolumn{1}{r|}{$+ \mathcal{L}_{\text{VGG}}$} & 0.1000 & 3.256 & 31.86 & 0.9001\\
        \multicolumn{1}{r|}{$+ \mathcal{L}_{\text{VGG}} + \mathcal{L}_{\text{Reblur, n2}}$} & \textbf{0.0868} & \textbf{2.835} & 31.83 & 0.9015\\
        \multicolumn{1}{r|}{$+ \mathcal{L}_{\text{Adv}}$} & 0.1158 & 3.395 & 31.84 & 0.8993\\
        \multicolumn{1}{r|}{$+ \mathcal{L}_{\text{Adv}} + \mathcal{L}_{\text{Reblur, n2}}$} & \textbf{0.0934} & \textbf{2.836} & 32.00 & 0.9061\\    % result from a very interesting learning curve: discriminator began to win from the 60% epoch of the training
        % \multicolumn{1}{r|}{$+ \mathcal{L}_{\text{Adv}} + \mathcal{L}_{\text{Reblur, n2}}$} & 0.1903 & 2.899 & 28.62 & 0.8592\\
        \midrule
        DHN ($\mathcal{L}_{1}$ only) & 0.0942 & 3.288 & 32.65 & 0.9152\\
        \multicolumn{1}{r|}{$+ \mathcal{L}_{\text{VGG}}$} & 0.0812 & 3.171 & 32.61 & 0.9146\\
        \multicolumn{1}{r|}{$+ \mathcal{L}_{\text{VGG}} + \mathcal{L}_{\text{Reblur, n2}}$} & \textbf{0.0723} & \textbf{2.821} & 32.48 & 0.9133\\
        \multicolumn{1}{r|}{$+ \mathcal{L}_{\text{Adv}}$} & 0.0956 & 3.218 & 32.58 & 0.9128\\
        \multicolumn{1}{r|}{$+ \mathcal{L}_{\text{Adv}} + \mathcal{L}_{\text{Reblur, n2}}$} & \textbf{0.0820} & \textbf{2.809} & 32.45 & 0.9121\\
        \bottomrule
    \end{tabularx}
    \\
    \tabcspace
    \caption{
        % \textbf{Results on REDS~\cite{Nah_2019_CVPR_Workshops_REDS} dataset from combined perceptual losses.}
        \textbf{Results on REDS dataset by adding reblurring loss to the other preceptual losses.}
        % Coefficients are omitted for better readability.
    }
    \label{tab:reds_vgg_reblur}
    \tabspace
\end{table}

Our reblurring loss is a new perceptual loss that is sensitive to blurriness of an image, a type of image structure-level information while other perceptual losses such as VGG loss~\citep{Johnson2016Perceptual} and adversarial loss~\citep{Ledig_2017_CVPR} are more related to the high-level contexts.
As VGG model~\citep{simonyan2014very} is trained to recognize image classes, optimizing with VGG loss could make an image better recognizable.
In the GAN frameworks~\citep{goodfellow2014generative}, it is well known that discriminators can easily tell fake images from real images~\citep{Wang_2020_CVPR_easy}, being robust against JPEG compression and blurring.
In the adversarial loss from the discriminator, the realism difference could be more salient than other features such as blurriness. 

With the perceptual loss functions designed with different objectives, combining them could bring visual quality improvements in various aspects.
Tables~\ref{tab:gopro_vgg_reblur} and \ref{tab:reds_vgg_reblur} show the effect of applying our reblurring loss jointly with other perceptual losses on GOPRO and REDS datasets.
We omit the loss coefficients for simplicity.
We used weight $0.3$ for the VGG loss $\mathcal{L}_{\text{VGG}}$ and $0.001$ for the adversarial loss, $\mathcal{L}_{\text{Adv}}$.
We witness LPIPS and NIQE further improves when our reblurring loss is combined with $\mathcal{L}_{\text{VGG}}$ or $\mathcal{L}_{\text{Adv}}$.

\section{Test-Time Adaptation Details}

We describe the detailed self-supervised test-time adaptation process.
At test time, the learning rate is set to $\mu=3\times10^{-6}$.
From the initial deblurring result $L^{0}$, the self-supervised loss is iteratively minimized by updating the weights of $\mathcal{M}_{\text{D}}$.
As the self-supervised loss in \eqref{eq:eq_loss_reblur_self} only cares about image sharpness, the image may have color drifting issues.
Thus, finally, we match the histogram of the updated image $L^N$ to the histogram of $L^0$.
The overall process is shown in Algorithm~\ref{alg:tta}.

\input{sections/algorithm/tta_alg}

\section{Perception vs. Distortion Trade-Off}
\label{sec:trade_off}

It is known in image restoration literature that the distortion error and the perceptual quality error are in trade-off relation~\citep{blau2018perception,Blau_2018_ECCV_Workshops}.
The relation is often witnessed by training a single model with different loss functions.
In most cases, to obtain a better perceptual quality from a single model architecture, retraining with another loss from scratch is necessary.
Our test-time adaptation from self-supervised reblurring loss, in contrast, can provide the steps toward perceptual quality without full retraining.

\begin{figure}[h]
    \renewcommand{\wp}{0.47 \linewidth}
    \subfloat[PSNR vs LPIPS\label{fig:supp_tta_srn_psnr_lpips}]{\includegraphics[width=\wp]{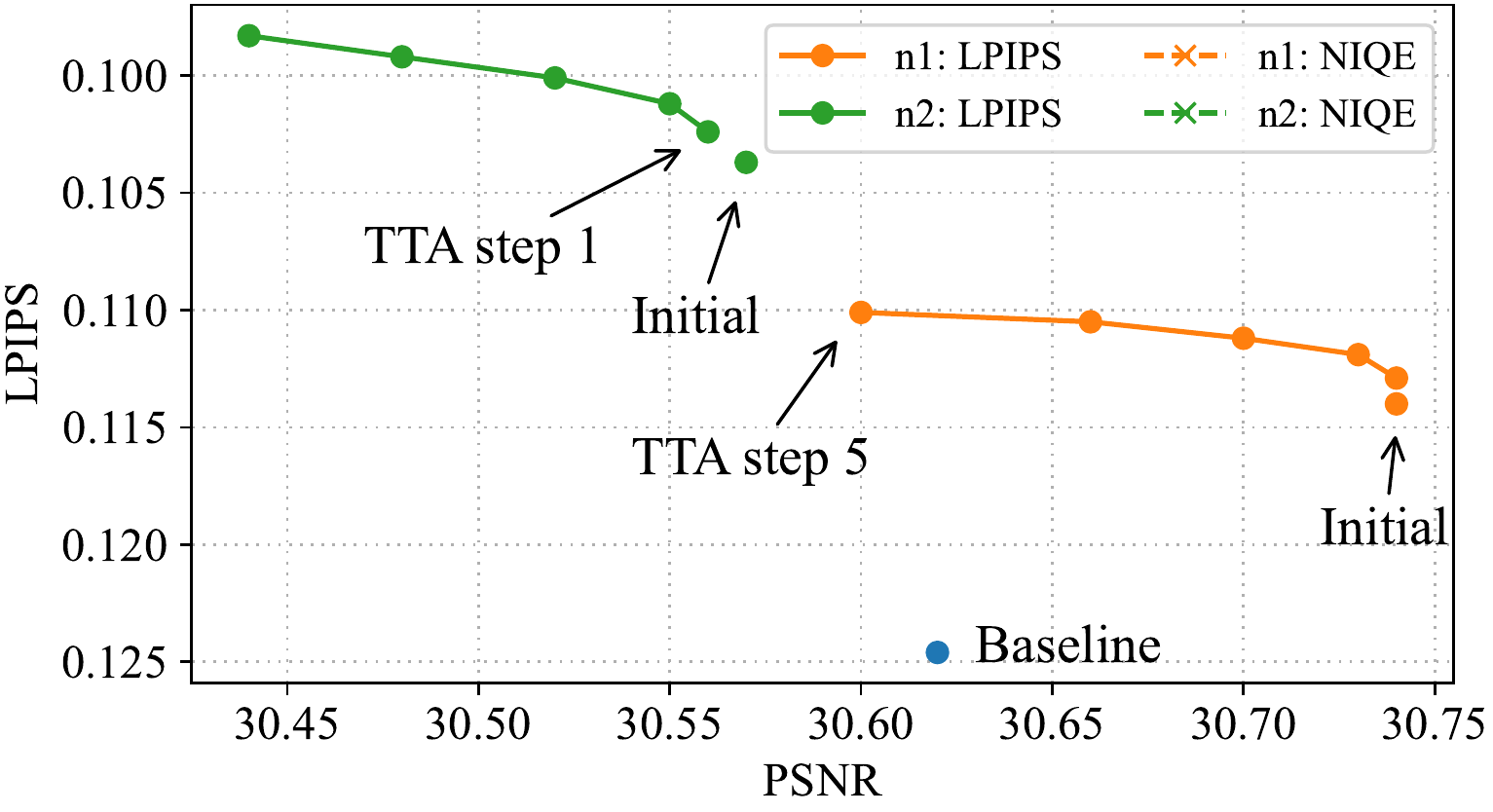}}
    \hfill
    \subfloat[PSNR vs NIQE\label{fig:supp_tta_srn_psnr_niqe}]{\includegraphics[width=\wp]{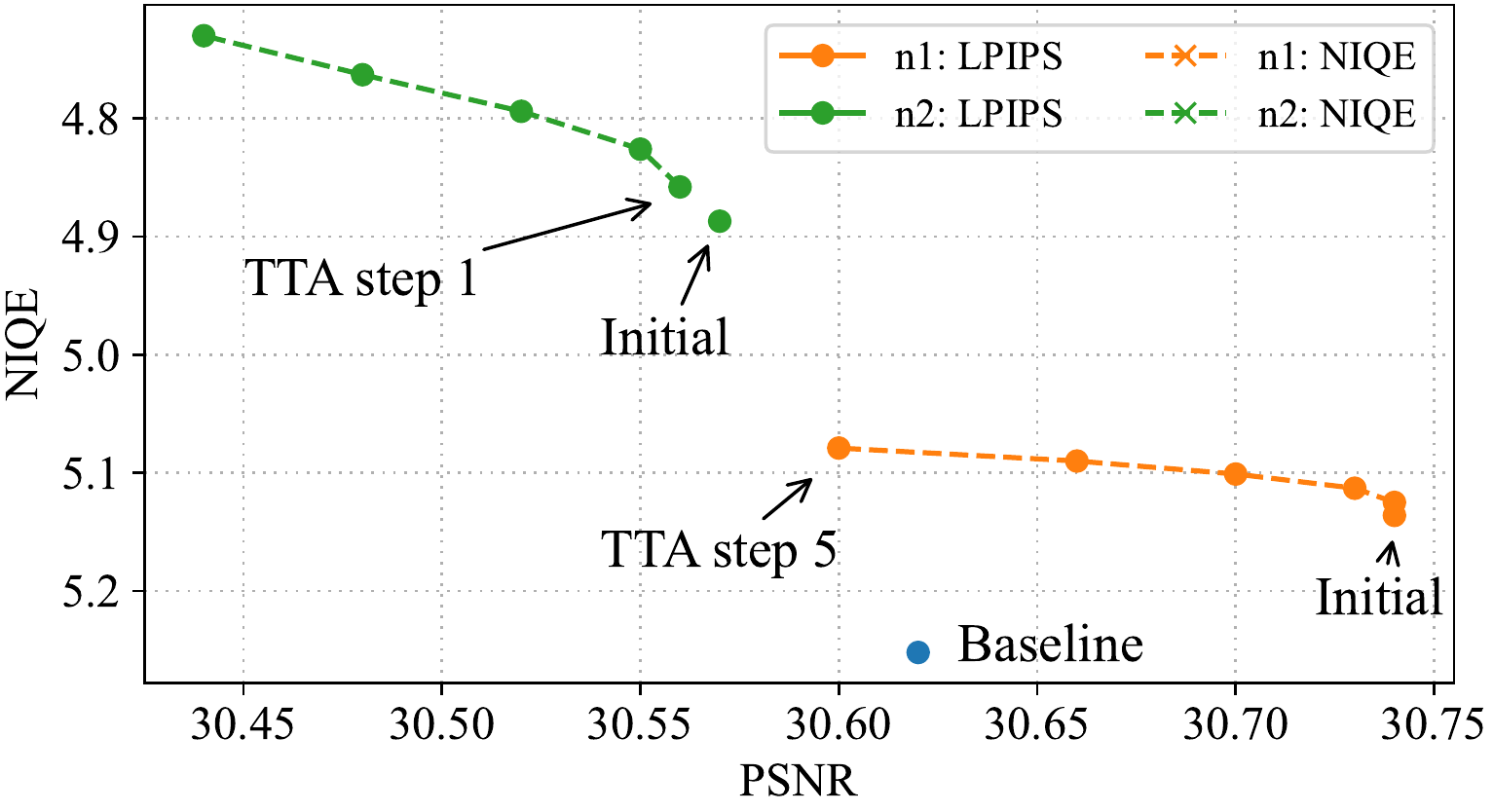}}
    \\
    \subfloat[SSIM vs LPIPS\label{fig:supp_tta_srn_ssim_lpips}]{\includegraphics[width=\wp]{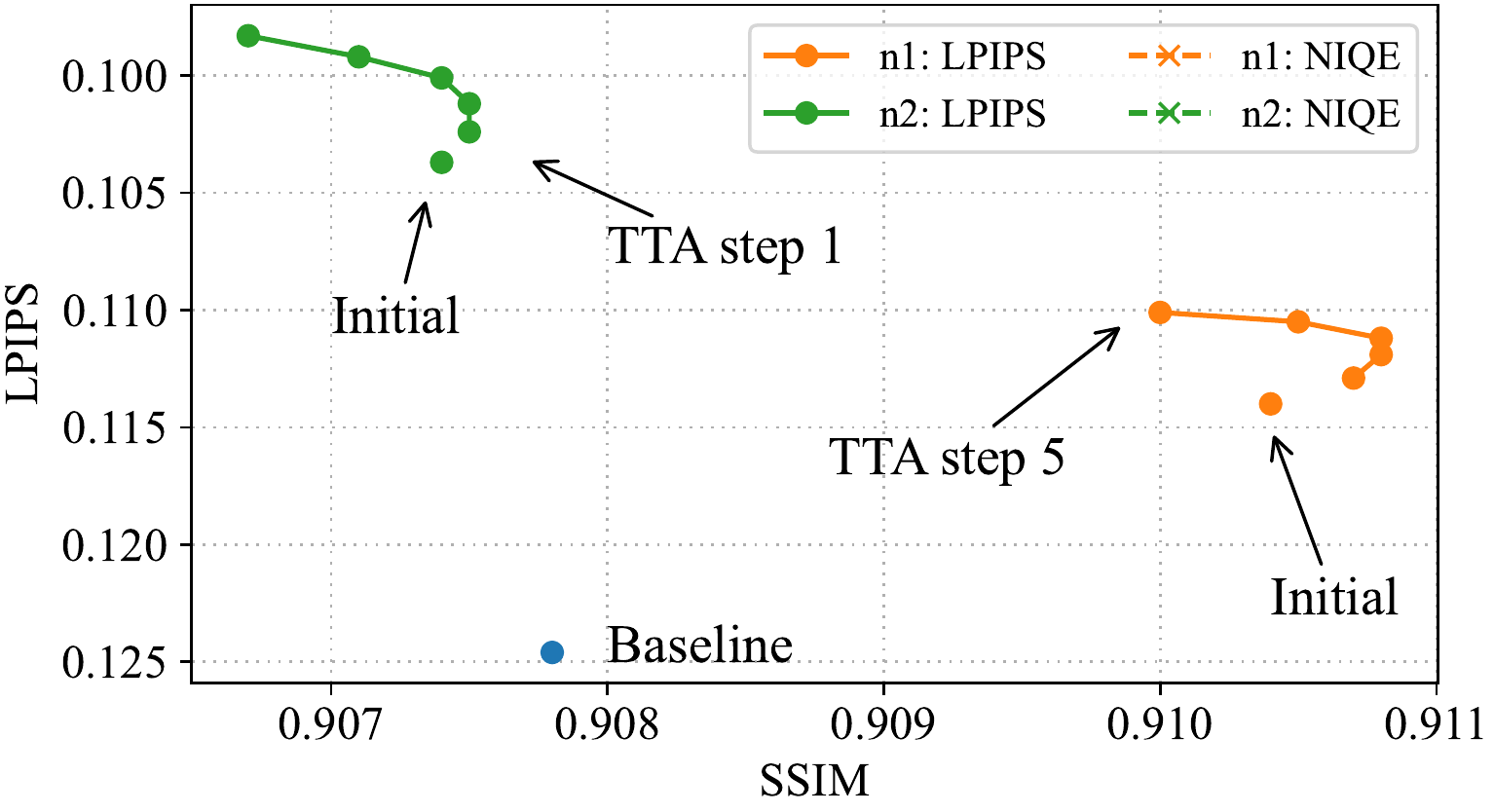}}
    \hfill
    \subfloat[SSIM vs NIQE\label{fig:supp_tta_srn_ssim_niqe}]{\includegraphics[width=\wp]{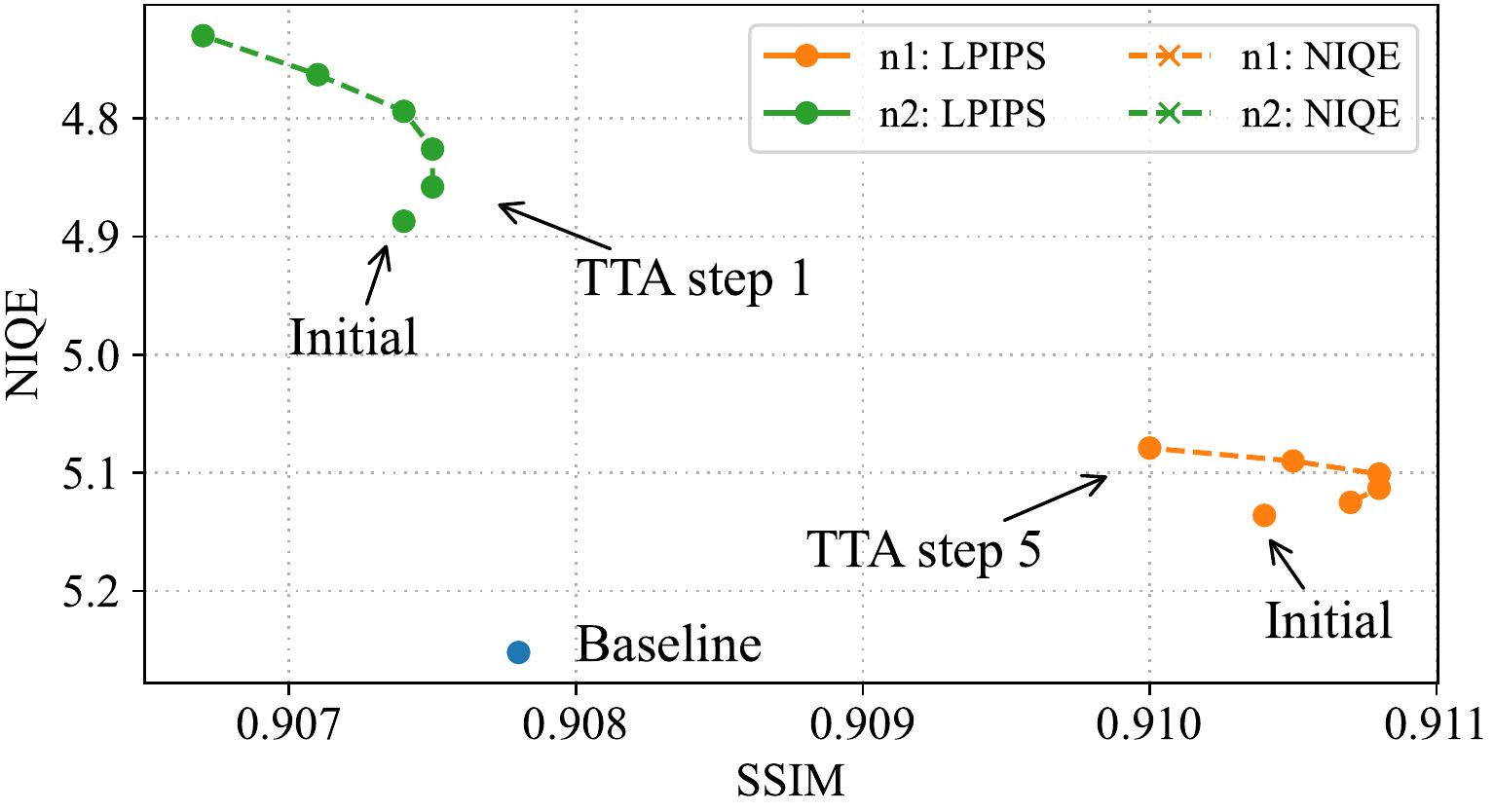}}
    \\
    \caption{
        \textbf{Perception-distortion trade-off from test-time adaptation applied to SRN model on GOPRO dataset.}
    }
    \label{fig:supp_tta_srn_trade_off}
    % \figspace
\end{figure}

\begin{figure}[h]
    \renewcommand{\wp}{0.47 \linewidth}
    \subfloat[PSNR vs LPIPS\label{fig:supp_tta_dhn_psnr_lpips}]{\includegraphics[width=\wp]{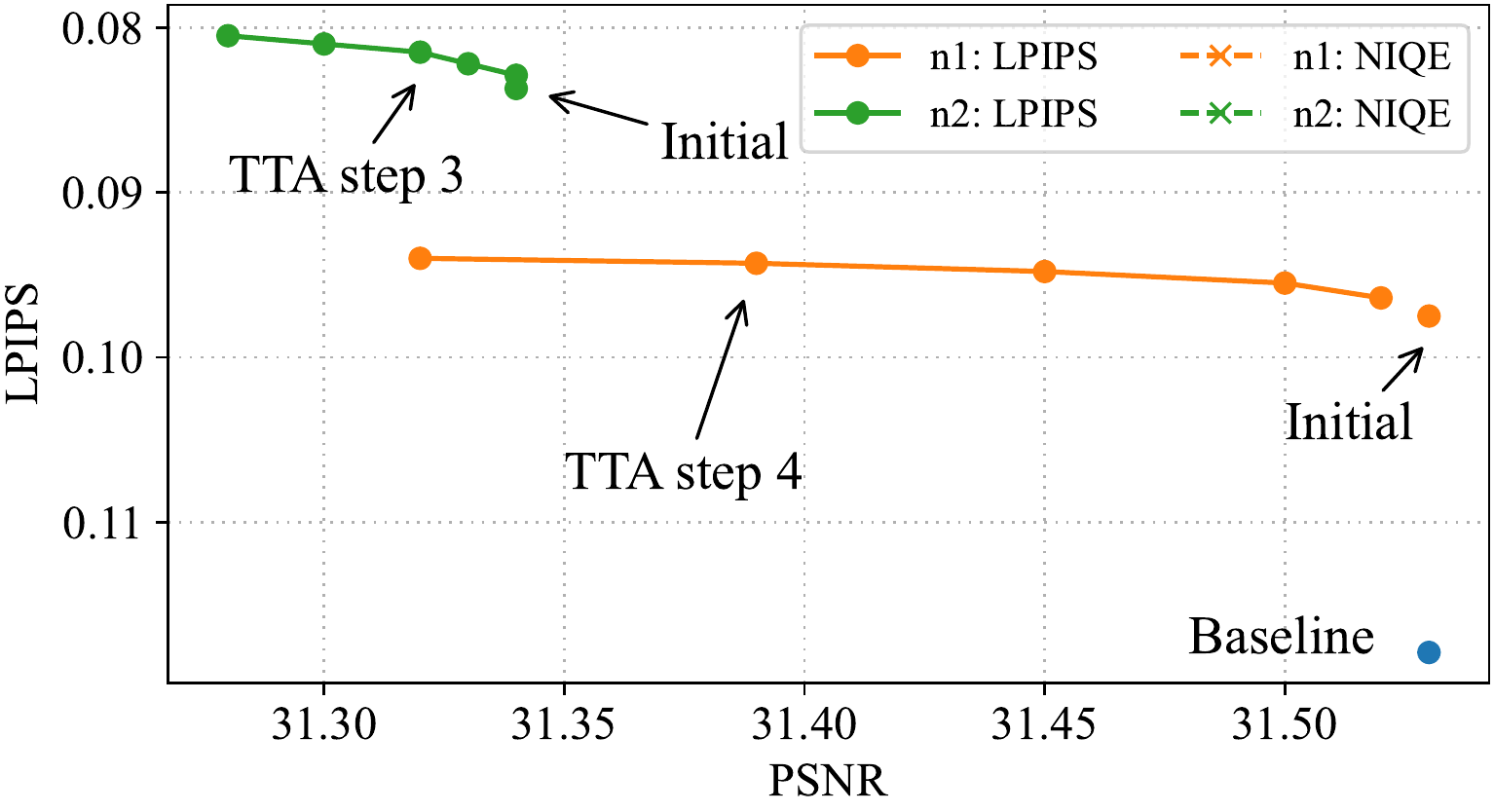}}
    \hfill
    \subfloat[PSNR vs NIQE\label{fig:supp_tta_dhn_psnr_niqe}]{\includegraphics[width=\wp]{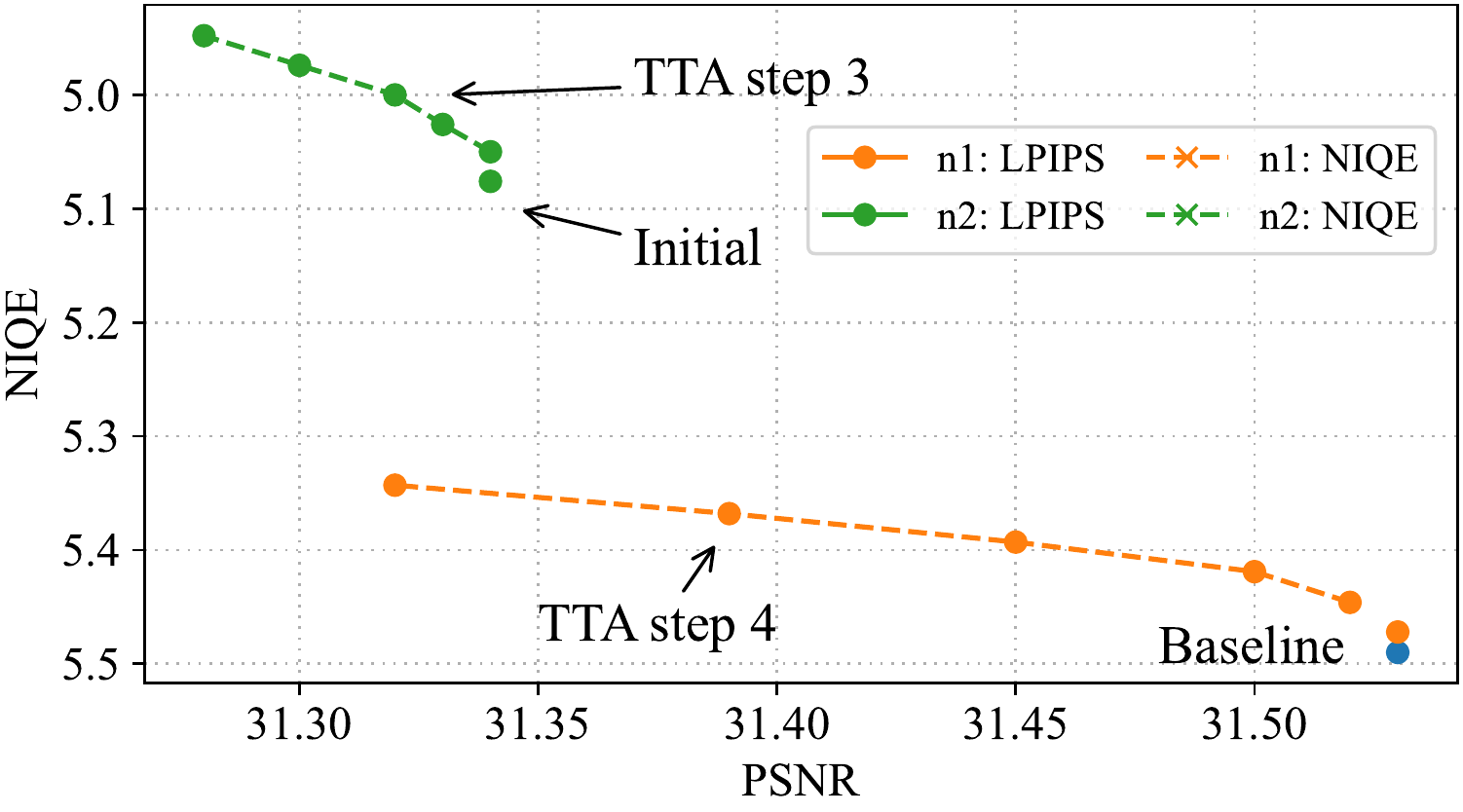}}
    \\
    \subfloat[SSIM vs LPIPS\label{fig:supp_tta_dhn_ssim_lpips}]{\includegraphics[width=\wp]{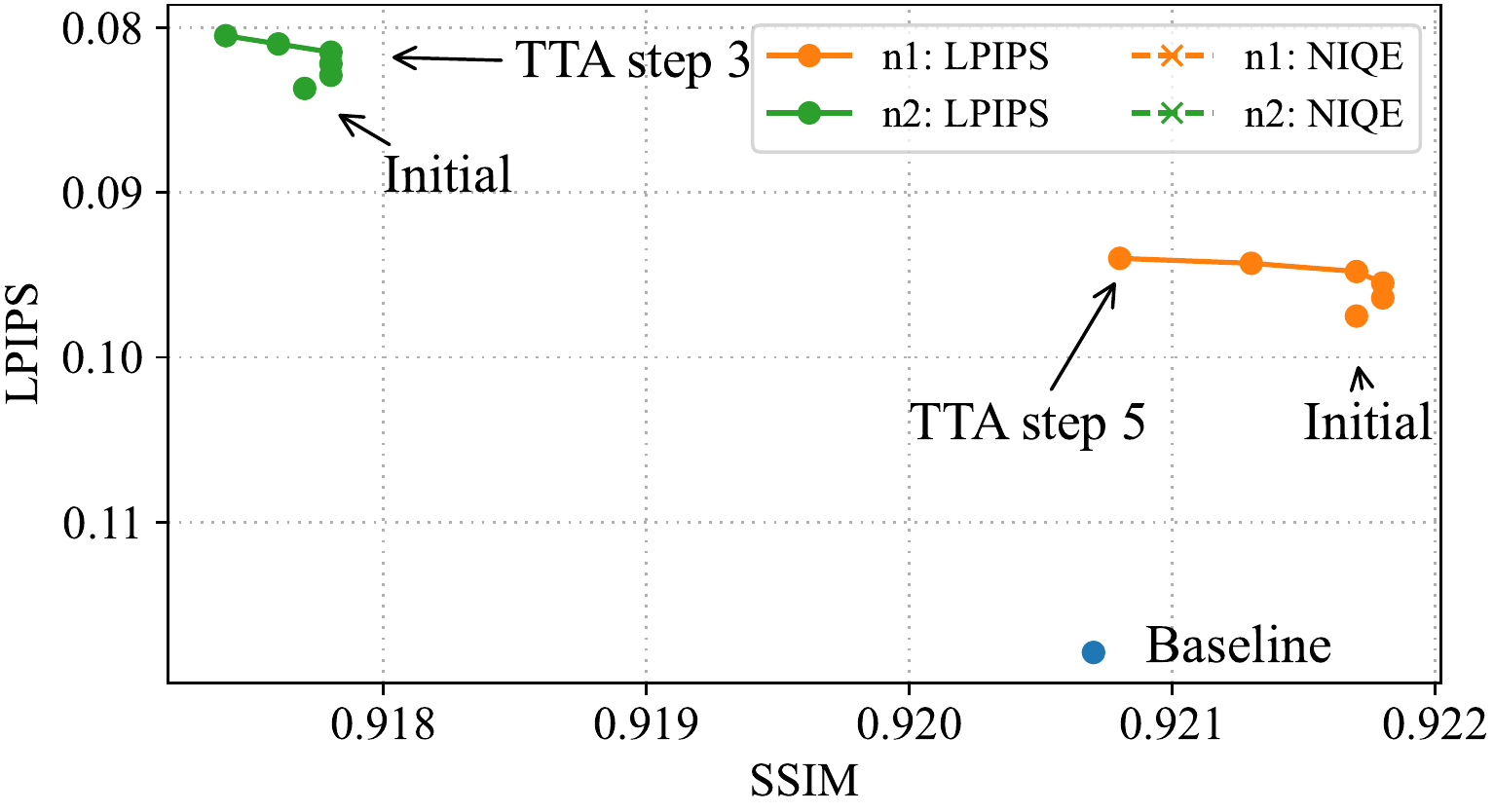}}
    \hfill
    \subfloat[SSIM vs NIQE\label{fig:supp_tta_dhn_ssim_niqe}]{\includegraphics[width=\wp]{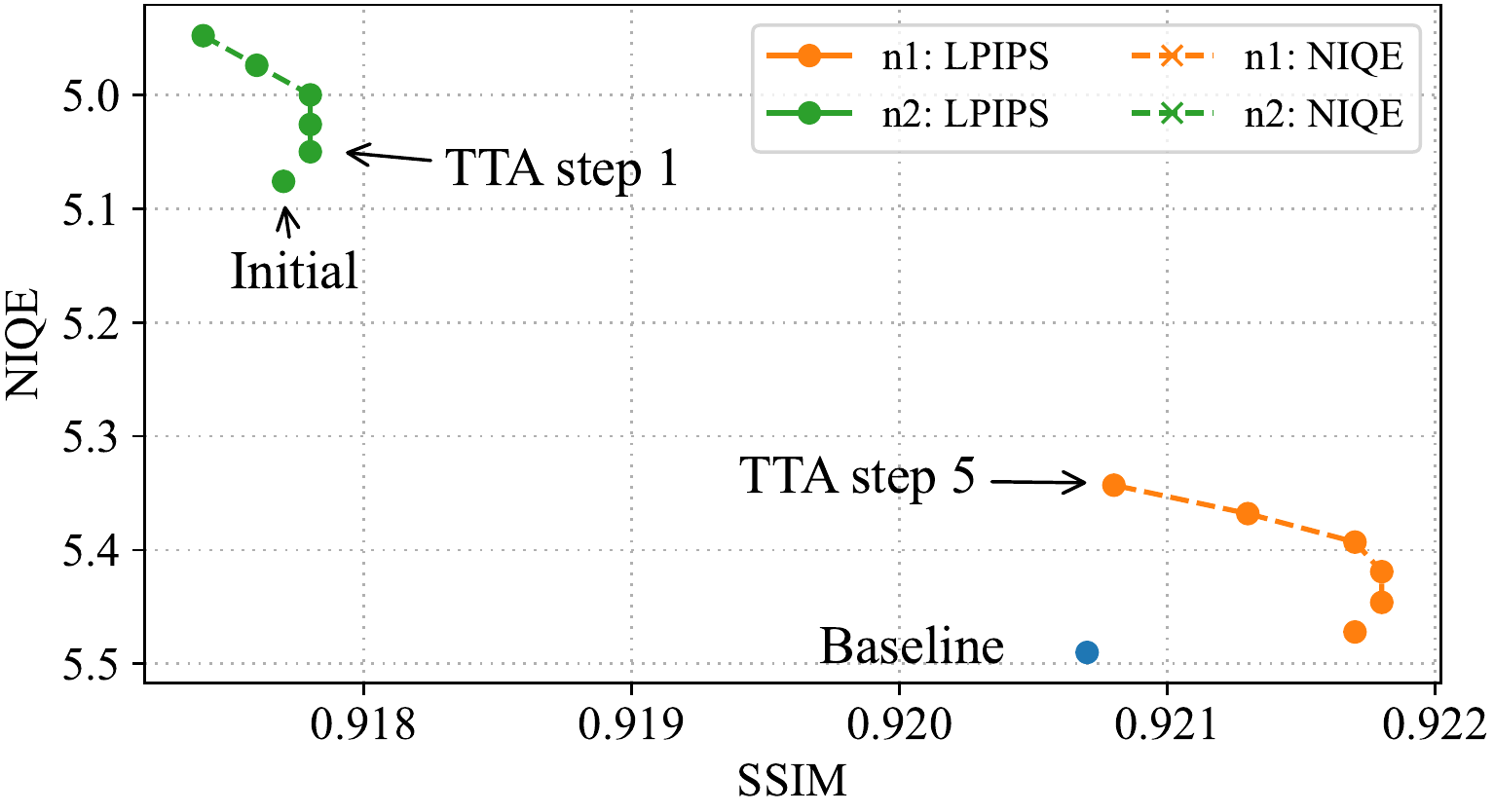}}
    \\
    % \figcspace
    \caption{
        \textbf{Perception-distortion trade-off from test-time adaptation applied to DHN model on GOPRO dataset.}
    }
    \label{fig:supp_tta_dhn_trade_off}
    % \figspace
\end{figure}

In Figure~\ref{fig:supp_tta_srn_trade_off} and \ref{fig:supp_tta_dhn_trade_off}, we present the perception-distortion trade-off from our test-time adaption.
LPIPS and NIQE scores consistently improve from each adaptation step in both SRN and DHN models.
While PSNR is moderately sacrificed from the adaptation, SSIM improves in the early steps as it more reflects the structural information.
Our results show improved trade-off between the distortion and perception metrics over the baseline models trained with L1 loss.

\iffalse

\section{Visual Comparison of Loss Function}
\label{sec:more_examples}
In this section, we present the visual comparison of deblurred results.
In Figure~\ref{fig:fig_ablation2} and \ref{fig:fig_ablation3}, we perform visual ablation by showing the deblurred results from baseline L1 loss, our reblurring loss, and additional test-time adaptation.
For $\mathcal{M}_{\text{D}}$, we used DHN.
For $\mathcal{M}_{\text{R}}$, 2 ResBlocks are used.
Our final result reveals sharper image structure and texture.

In Figure~\ref{fig:fig_ablation4} and \ref{fig:fig_ablation5}, we compare the effect of 3 different perceptual losses.
The results from VGG loss, adversarial loss, and our reblurring loss are shown.
Our reblurring loss exhibits clearer edges and the face details than other perceptual losses.

\fi

\section{Test-Time Adaptation Effects}

In Table~\ref{tab:gopro_reds_tta_n}, we quantitatively compare the deblurred results from test-time adaptation in terms of a no-reference metric, NIQE, and reference-based metrics, LPIPS, PSNR, and SSIMs.
By performing TTA up to 30 steps as described in Algorithm~\ref{alg:tta}, we show that LPIPS and NIQE could be improved to a degree.
On both GOPRO and REDS datasets, NIQE has a tendency to improve further after LPIPS has stopped its improvement.
This is due to the self-supervised nature of test-time adaptation that considers image sharpness without reference.

\begin{table}[!t]
    \centering
    \footnotesize
    \renewcommand{\arraystretch}{1.15}
    \newcommand{\justlone}{\multicolumn{1}{l |}{$\mathcal{L}_{1}$}}
    \newcommand{\plusreblur}[1]{\multicolumn{1}{l |}{$\mathcal{L}_{1} + \mathcal{L}_{\text{Reblur, n{#1}}}$}}
    
    \begin{tabularx}{\linewidth}{l | p{2.2cm} >{\centering\arraybackslash}X >{\centering\arraybackslash}X >{\centering\arraybackslash}X >{\centering\arraybackslash}X >{\centering\arraybackslash}X >{\centering\arraybackslash}X >{\centering\arraybackslash}X >{\centering\arraybackslash}X}
        \toprule
        &  & \multicolumn{4}{c}{On GOPRO dataset} & \multicolumn{4}{c}{On REDS dataset} \\
        Model & \multicolumn{1}{c |}{Optimization} & LPIPS$_\downarrow$ & NIQE$_\downarrow$ & PSNR$^\uparrow$ & \multicolumn{1}{c |}{SSIM$^\uparrow$} &  LPIPS$_\downarrow$ & NIQE$_\downarrow$ & PSNR$^\uparrow$ & SSIM$^\uparrow$ \\
        \midrule
        \multirow{11}{*}{SRN} & \justlone & 0.1246 & 5.252 & 30.62 & \multicolumn{1}{c |}{0.9078} & 0.1148 & 3.392 & 31.89 & 0.8999 \\
        % \midrule
        & \plusreblur{1} & 0.1140 & 5.136 & 30.74 & \multicolumn{1}{c |}{0.9104} & 0.1071 & 3.305 & 32.01 & 0.9044 \\
        & \multicolumn{1}{r |}{$+$ TTA step \phantom{0}5} & \textbf{0.1101} & 5.079 & 30.60 & \multicolumn{1}{c |}{0.9100} & 0.1029 & 3.278 & 31.83 & 0.9040 \\
        & \multicolumn{1}{r |}{$+$ TTA step 10} & 0.1103 & 5.036 & 30.11 & \multicolumn{1}{c |}{0.9048} & \textbf{0.1025} & \textbf{3.261} & 31.29 & 0.8996 \\
        & \multicolumn{1}{r |}{$+$ TTA step 20} & 0.1223 & 4.968 & 28.44 & \multicolumn{1}{c |}{0.8806} & 0.1116 & 3.265 & 29.59 & 0.8807 \\
        & \multicolumn{1}{r |}{$+$ TTA step 30} & 0.1470 & \textbf{4.924} & 26.42 & \multicolumn{1}{c |}{0.8411} & 0.1306 & 3.301 & 27.73 & 0.8523 \\
        % \midrule
        & \plusreblur{2} & 0.1037 & 4.887 & 30.57 & \multicolumn{1}{c |}{0.9074} & 0.0947 & 2.875 & 31.82 & 0.9026 \\
        & \multicolumn{1}{r |}{$+$ TTA step \phantom{0}5} & 0.0983 & 4.730 & 30.44 & \multicolumn{1}{c |}{0.9067} & \textbf{0.0909} & 2.798 & 31.50 & 0.9008 \\
        & \multicolumn{1}{r |}{$+$ TTA step 10} & \textbf{0.0962} & 4.569 & 30.07 & \multicolumn{1}{c |}{0.9024} & 0.0913 & 2.741 & 30.87 & 0.8945 \\
        & \multicolumn{1}{r |}{$+$ TTA step 20} & 0.1021 & 4.274 & 28.83 & \multicolumn{1}{c |}{0.8836} & 0.1033 & \textbf{2.699} & 29.09 & 0.8697 \\
        & \multicolumn{1}{r |}{$+$ TTA step 30} & 0.1199 & \textbf{4.045} & 27.26 & \multicolumn{1}{c |}{0.8529} & 0.1259 & 2.729 & 27.20 & 0.8326 \\
        \bottomrule
    \end{tabularx}
    \\
    % \tabcspace
    \caption{
        \textbf{Quantitative analysis of the reblurring losses and test-time adaptation applied to SRN on GOPRO and REDS datasets.}
        }
    \label{tab:gopro_reds_tta_n}
    % \tabspace
\end{table}

\begin{figure}[t]
    \newcommand{\fwidth}{0.495 \linewidth}
    \newcommand{\pwidth}{0.245 \linewidth}
    \centering
    \subfloat[Blur]{\includegraphics[width=\fwidth]{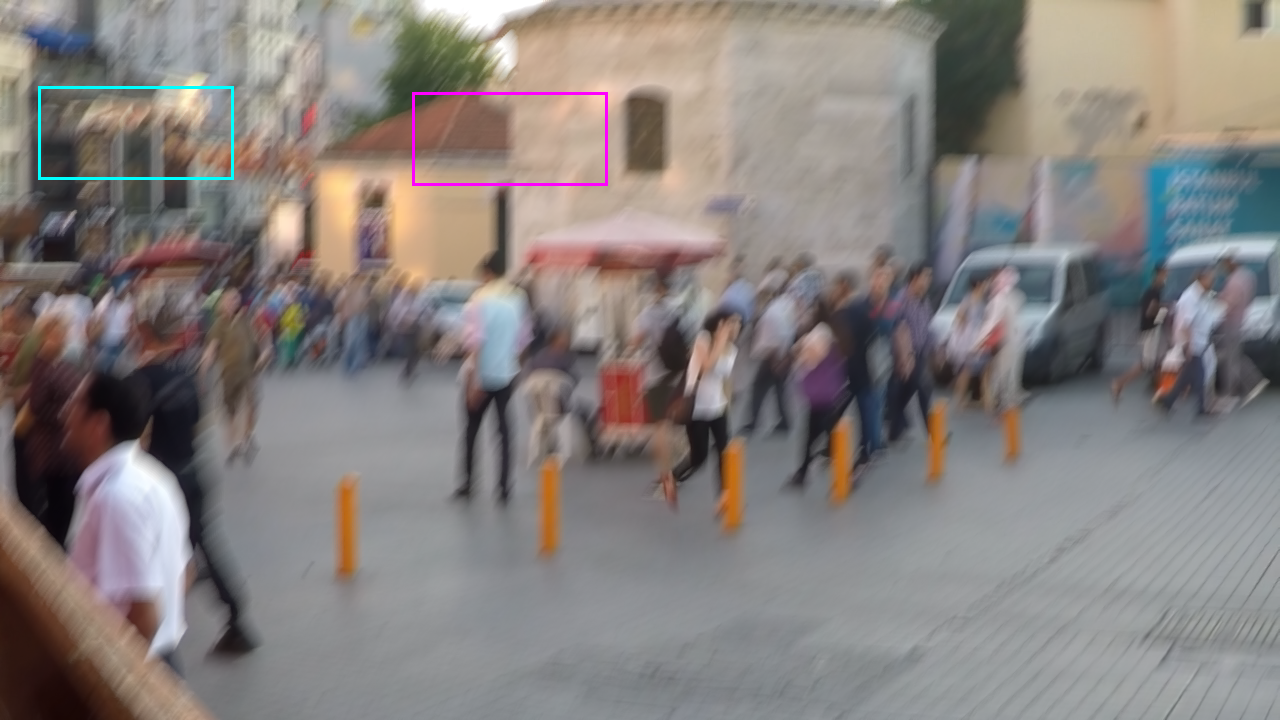}}
    \hfill
    \subfloat[Our deblurred image (TTA step 30)]{\includegraphics[width=\fwidth]{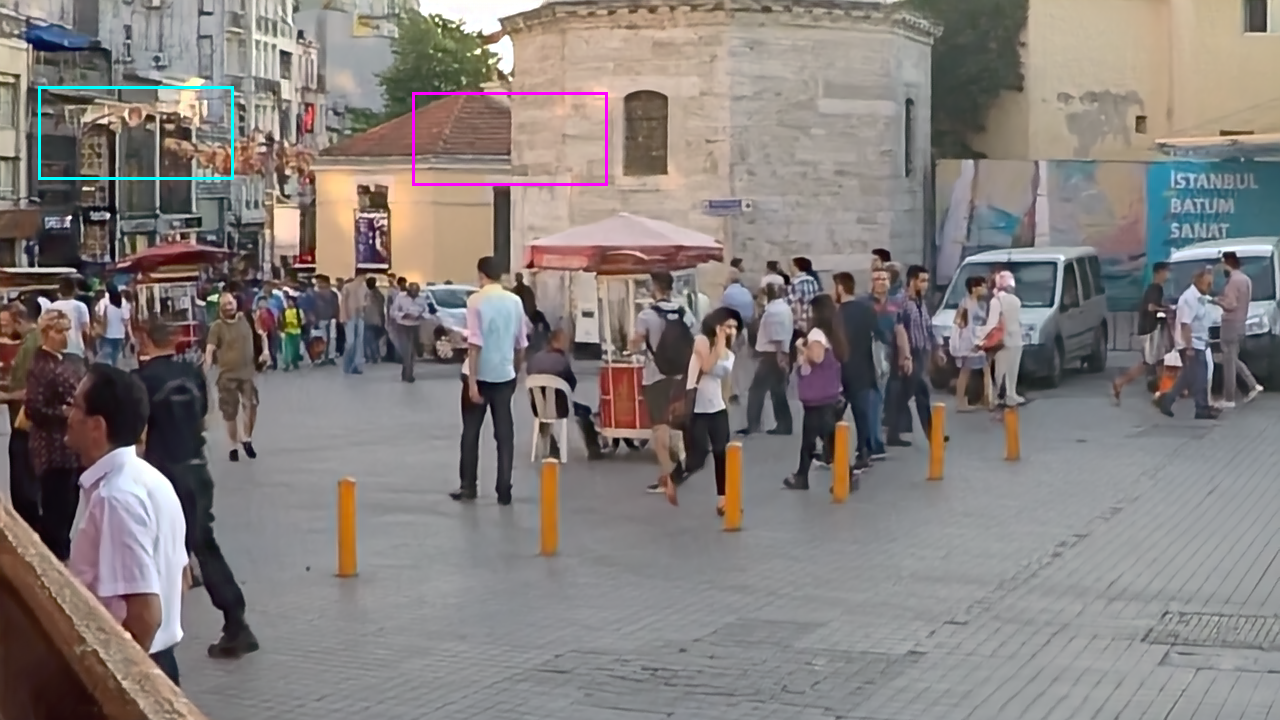}}
    \\
    \subfloat{\includegraphics[width=\pwidth]{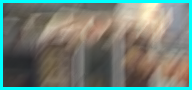}}
    \hfill
    \subfloat{\includegraphics[width=\pwidth]{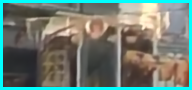}}
    \hfill
    \subfloat{\includegraphics[width=\pwidth]{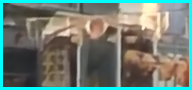}}
    \hfill
    \subfloat{\includegraphics[width=\pwidth]{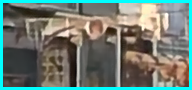}}
    \\
    \addtocounter{subfigure}{-4}
    \subfloat[Blur $B$]{\includegraphics[width=\pwidth]{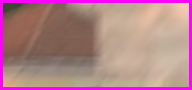}}
    \hfill
    \subfloat[$\mathcal{L}_1$]{\includegraphics[width=\pwidth]{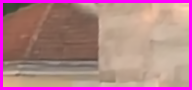}}
    \hfill
    \subfloat[$\mathcal{L}_1 + \mathcal{L}_{\text{Reblur, n1}}$]{\includegraphics[width=\pwidth]{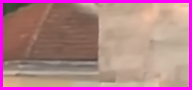}}
    \hfill
    \subfloat[Ours (TTA step 30)]{\includegraphics[width=\pwidth]{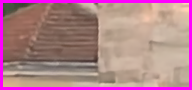}}
    \\
    \figcspace
    \caption{
        \textbf{Visual comparison of deblurred results by reblurring loss and test-time adaptation on GOPRO dataset.
        }
    }
    \label{fig:gopro_tta_n}
    \figspace
\end{figure}

\begin{figure}[t]
    \newcommand{\fwidth}{0.495 \linewidth}
    \newcommand{\pwidth}{0.245 \linewidth}
    \centering
    \subfloat[Blur]{\includegraphics[width=\fwidth]{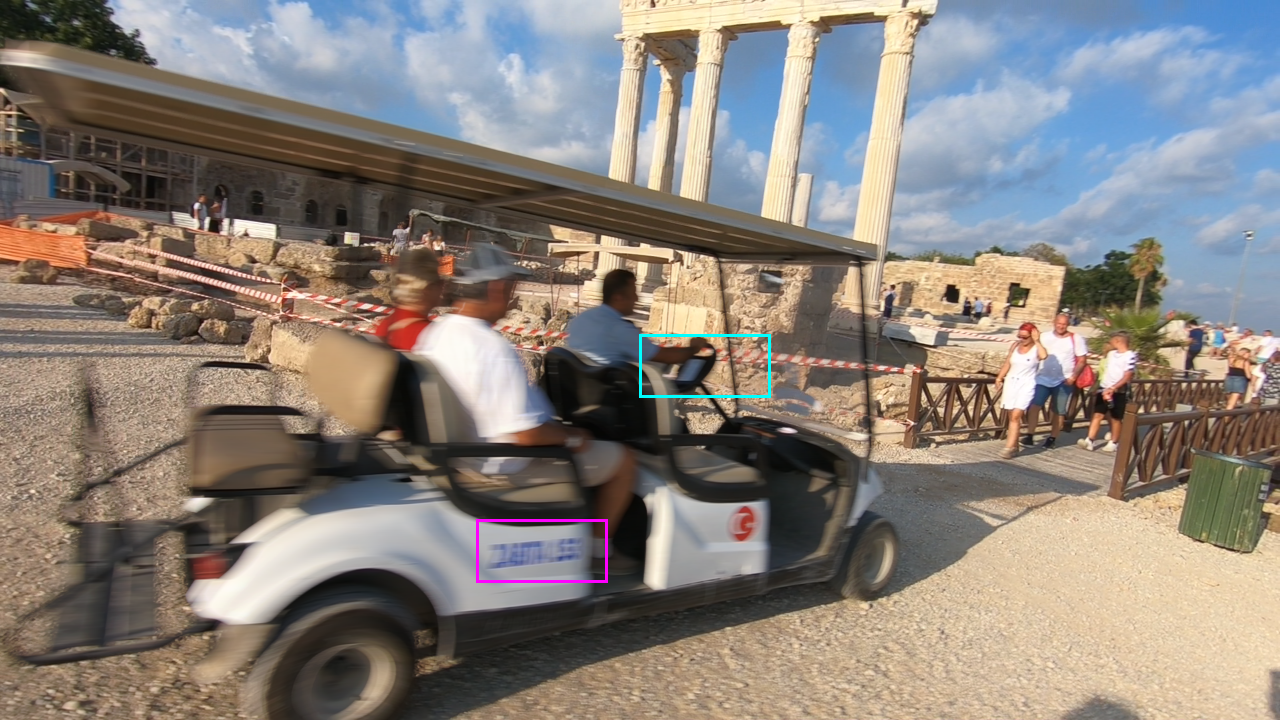}}
    \hfill
    \subfloat[Our deblurred image (TTA step 30)]{\includegraphics[width=\fwidth]{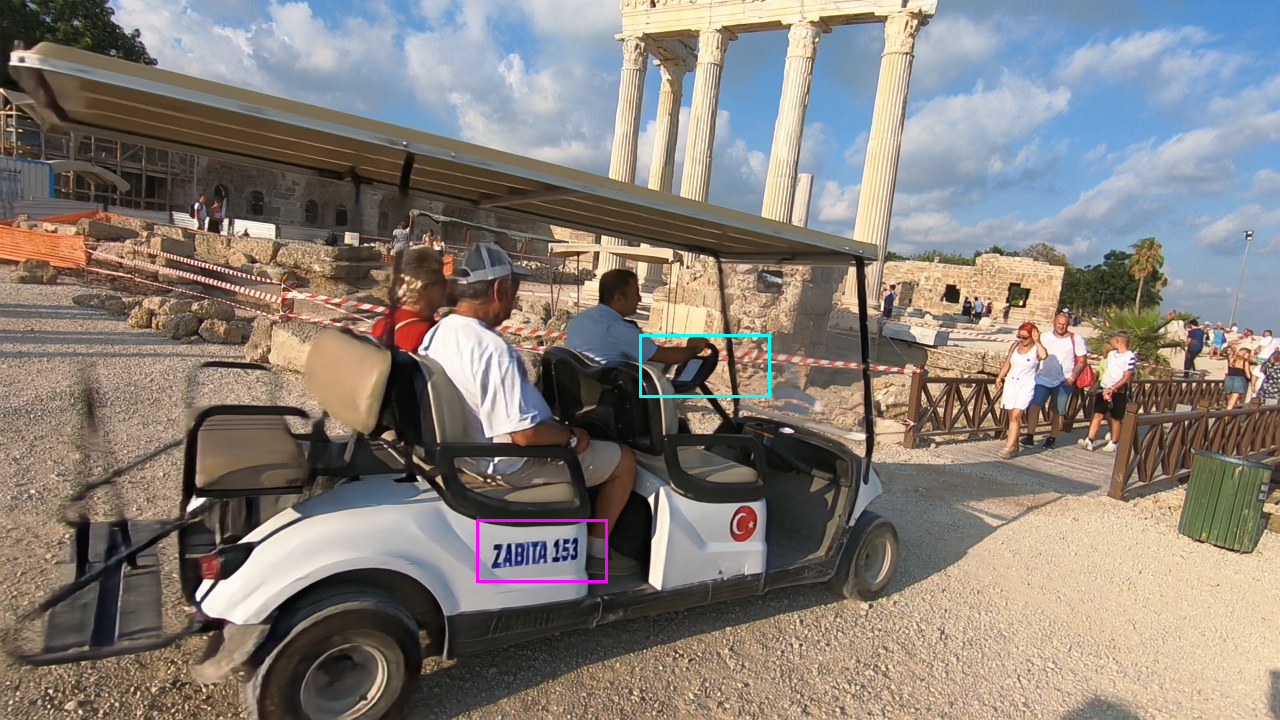}}
    \\
    \subfloat{\includegraphics[width=\pwidth]{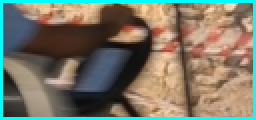}}
    \hfill
    \subfloat{\includegraphics[width=\pwidth]{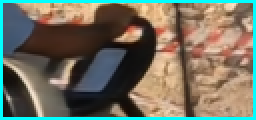}}
    \hfill
    \subfloat{\includegraphics[width=\pwidth]{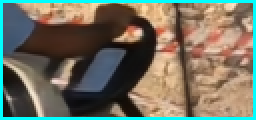}}
    \hfill
    \subfloat{\includegraphics[width=\pwidth]{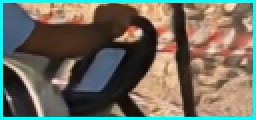}}
    \\
    \addtocounter{subfigure}{-4}
    \subfloat[Blur $B$]{\includegraphics[width=\pwidth]{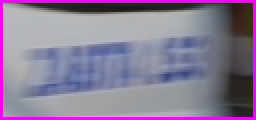}}
    \hfill
    \subfloat[$\mathcal{L}_1$]{\includegraphics[width=\pwidth]{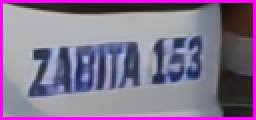}}
    \hfill
    \subfloat[$\mathcal{L}_1 + \mathcal{L}_{\text{Reblur, n1}}$]{\includegraphics[width=\pwidth]{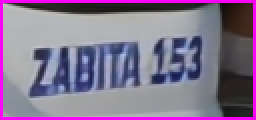}}
    \hfill
    \subfloat[Ours (TTA step 30)]{\includegraphics[width=\pwidth]{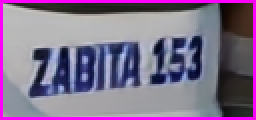}}
    \\
    \figcspace
    \caption{
        \textbf{Visual comparison of deblurred results by reblurring loss and test-time adaptation on REDS dataset.
        }
    }
    \label{fig:reds_tta_n}
    \figspace
\end{figure}

In Figures~\ref{fig:gopro_tta_n} and \ref{fig:reds_tta_n}, we visually show the effect of test-time adaptation applied to SRN with a jointly trained reblurring module. 
By test-time adaptation, our model further improves the sharp edges of the images.
In Figure~\ref{fig:gopro_tta_n}, the building structures and the horizontal lines are better witnessed.
Also in Figure~\ref{fig:reds_tta_n}, the vehicle's pole are better recovered and the text are clearer.
While the PSNR and SSIM have decreased by test-time adaptation in Table~\ref{tab:gopro_reds_tta_n}, perceptually, the results from test-time adaptation tend to be sharper.

\clearpage

%% file: sections/algorithm/tta_alg.tex
\begin{algorithm}[h]
    \setstretch{1.10}
    \caption{Optimization process in test-time adaptation}
    \begin{algorithmic}[1]
        \Procedure{Test-time Adaptation}{$B, \mathcal{M}_{\text{D}}, \mathcal{M}_{\text{R}}$}
        \State Test-time learning rate $\mu \gets 3\times10^{-6}$.
        \State $\theta_{\text{D}} \gets$ Weights of $\mathcal{M}_{\text{D}}$.
        \State $L^{0} = \mathcal{M}_{\text{D}}(B)$.
        \iffalse
        \For{$i = 1 \dots N$}
            \State Fix $L^{i-1}$
            \State $\mathcal{L}_{\text{reblur}}^{\text{self}} = \|\mathcal{M}_{\text{R}}(\mathcal{M}_{\text{D}}(B)) - {L}^{i-1}\|$.
            \State update $\theta_{\text{D}}$ by learning rate $\mu$
            \State $L^{i} = \mathcal{M}_{\text{D}}(B)$
        \EndFor
        \fi
        \For{$i = 0 \dots N - 1$}
        \State $L^i_\ast = \mathcal{M}_\text{D} \paren{B}$.
        \State $\mathcal{L}_{\text{reblur}}^{\text{self}} = \|\mathcal{M}_{\text{R}}(\mathcal{M}_{\text{D}}(B)) - {L}^{i}_\ast\|$.
        \State Update $\theta_{\text{D}}$ by $\nabla_{\theta_\text{D}} \mathcal{L}_\text{Reblur}^\text{self}$ and $\mu$.
        \EndFor
        $L^N = \mathcal{M}_\text{D} \paren{ B }$.
        \State $L^N_\text{Adapted} = ${\fontfamily{qcr}\selectfont histogram\_matching}$(L^{N}, L^{0}_\ast)$
        \State \textbf{return} $L^N_\text{Adapted}$
        \EndProcedure
    \end{algorithmic}
    \label{alg:tta}
\end{algorithm}